\documentclass[twocolumn,superscriptaddress,article]{revtex4-1}
\usepackage{overpic}
\usepackage{graphicx, latexsym, verbatim}
\usepackage{graphics}
\usepackage{amssymb, amsmath}
\usepackage{color}
\usepackage{multirow}
\usepackage{epstopdf}
\usepackage{color}
\usepackage{multirow}
\definecolor{darkblue}{rgb}{0,0,0.55}
\definecolor{comfrey}{rgb}{0.85,0.85,0.85}
\usepackage{rotating}
\usepackage{pgfplots,tikz}
\newcommand{\ud}{\mathrm{d}}
\newcommand{\mr}{\mathbf{r}}

\newcommand{\mG}{\mathbf{G}}
\newcommand{\mk}{\mathbf{k}}
\newcommand{\mE}{\mathbf{E}}
\newcommand{\mH}{\mathbf{H}}
\newcommand{\mB}{\mathbf{B}}

\newcommand{\mf}{\mathbf{f}}
\newcommand{\mft}{\tilde{\mathbf{f}}}
\newcommand{\tlo}{\tilde{\omega}}
\newcommand{\ft}{\tilde{\text{f}}}
\newcommand{\f}{\text{f}}
\newcommand{\E}{\text{E}}

\newcommand{\tlk}{\tilde{k}}
\newcommand{\me}{\mathbf{e}}

\definecolor{myBlue}{rgb}{0.0430,0.5156,0.7773}

\newcommand{\comm}[1]{{\color{black}#1}}
\newcommand{\myComm}[1]{{\color{black}}}
\newcommand{\Jakob}[1]{{\color{black}#1}}
\newcommand{\jesm}[1]{{\color{black}#1}}
\newcommand{\ptk}[1]{{\color{black}#1}}
\newcommand{\ngre}[1]{{\color{black}#1}}

\newcommand{\rework}[1]{}
\newcommand{\takeout}[1]{{}}
\newcommand{\hmm}[1]{{\color{green}}}

\begin{document}


\title{On the theory of coupled modes in optical cavity-waveguide structures}
\author{Philip Tr{\o}st Kristensen}
\affiliation{Institut f\"ur Physik, Humboldt Universit\"at zu Berlin, 12489 Berlin, Germany}
\author{Jakob Rosenkrantz de Lasson}
\affiliation{DTU Fotonik, Technical University of Denmark, DK-2800 Kgs. Lyngby, Denmark}
\author{Mikkel Heuck}
\affiliation{DTU Fotonik, Technical University of Denmark, DK-2800 Kgs. Lyngby, Denmark}
\author{Niels Gregersen}
\affiliation{DTU Fotonik, Technical University of Denmark, DK-2800 Kgs. Lyngby, Denmark}
\author{Jesper M{\o}rk}
\affiliation{DTU Fotonik, Technical University of Denmark, DK-2800 Kgs. Lyngby, Denmark}

\date{\today}


\begin{abstract}
Light propagation in systems of optical cavities coupled to waveguides can be conveniently described by a general rate equation model known as (temporal) coupled mode theory (CMT). We present an alternative derivation of the CMT for optical cavity-waveguide structures, which explicitly relies on the treatment of the cavity modes as quasinormal modes with properties that are distinctly different from those of the modes in the waveguides. The two families of modes are coupled via the field equivalence principle to provide a physically appealing yet surprisingly accurate description of light propagation in the coupled systems. Practical application of the theory is illustrated using example calculations in one and two dimensions.
\end{abstract}


\maketitle

                                  \section{Introduction}
Coupled systems of optical waveguides and micro cavities provide a powerful platform for integrated optical components with applications ranging from optical experiments 
to communication networks. In experiments, the coupling to a waveguide provides convenient input and output channels for cavities~\cite{Cai_PRL_85_74_2000, Faraon_APL_90_073102_2007} in which the optical field is enhanced to increase light-matter interaction~\cite{Brossard_APL_97_111101_2010,Johnson_OE_14_817_2006}. For communication purposes, 
the micro cavities may act as 
filters to transmit or drop specific wavelengths~\cite{Noda_Nature_407_608_2000} or as sharp bends to guide the light in circuits with microscopic footprints~\cite{Lin_Science_282_274_1998}, and it has been shown that waveguides can act to couple distant cavities~\cite{Sato_NP_6_56_2012}. The relatively high optical energy density may lead to larger impact of non-linear material responses, such as the Kerr effect, and carrier and temperature induced index changes~\cite{Husko_APL_94_021111_2009, deRossi_PRA_79_043818_2009} which lead to shifts in the cavity resonance frequency and the cavity-waveguide coupling. This may, in turn, form the basis for optical buffers~\cite{Xu_NaturePhys_3_406_2007, Nozaki_NaturePhot_4_477_2010}  or integrated all-optical switching, in which control pulses of light are used to govern the transmission of signal pulses~\cite{Nozaki_NaturePhot_4_477_2010}, and promises ultra fast operation without the need for energy-consuming optical to electronic conversions. For ease in the interpretation of experiments and design of future integrated optical components, it is naturally of considerable interest to have both accurate and efficient theoretical models of light propagation in such coupled optical cavity-waveguide systems.
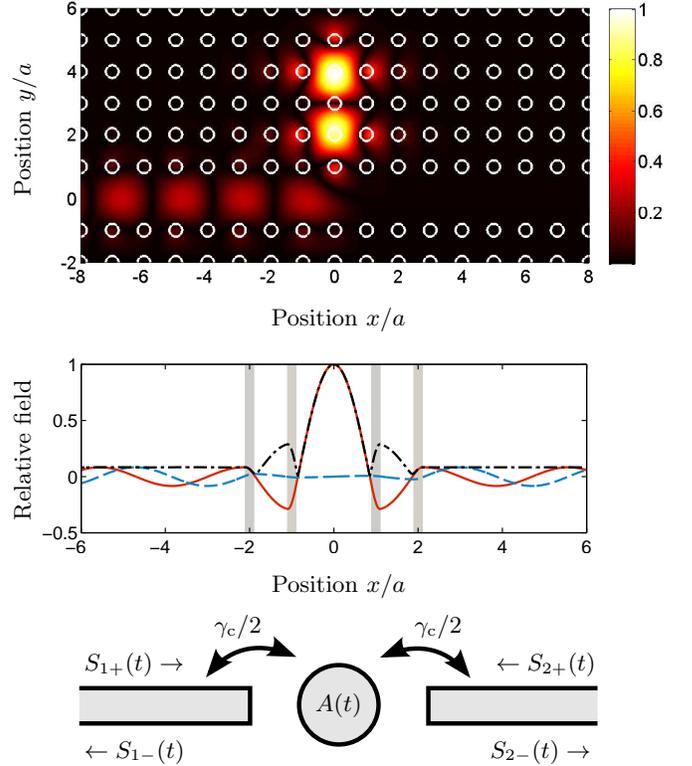
\begin{figure}[b!]
\flushright
\vspace{-3mm}
\begin{overpic}[width=8cm]{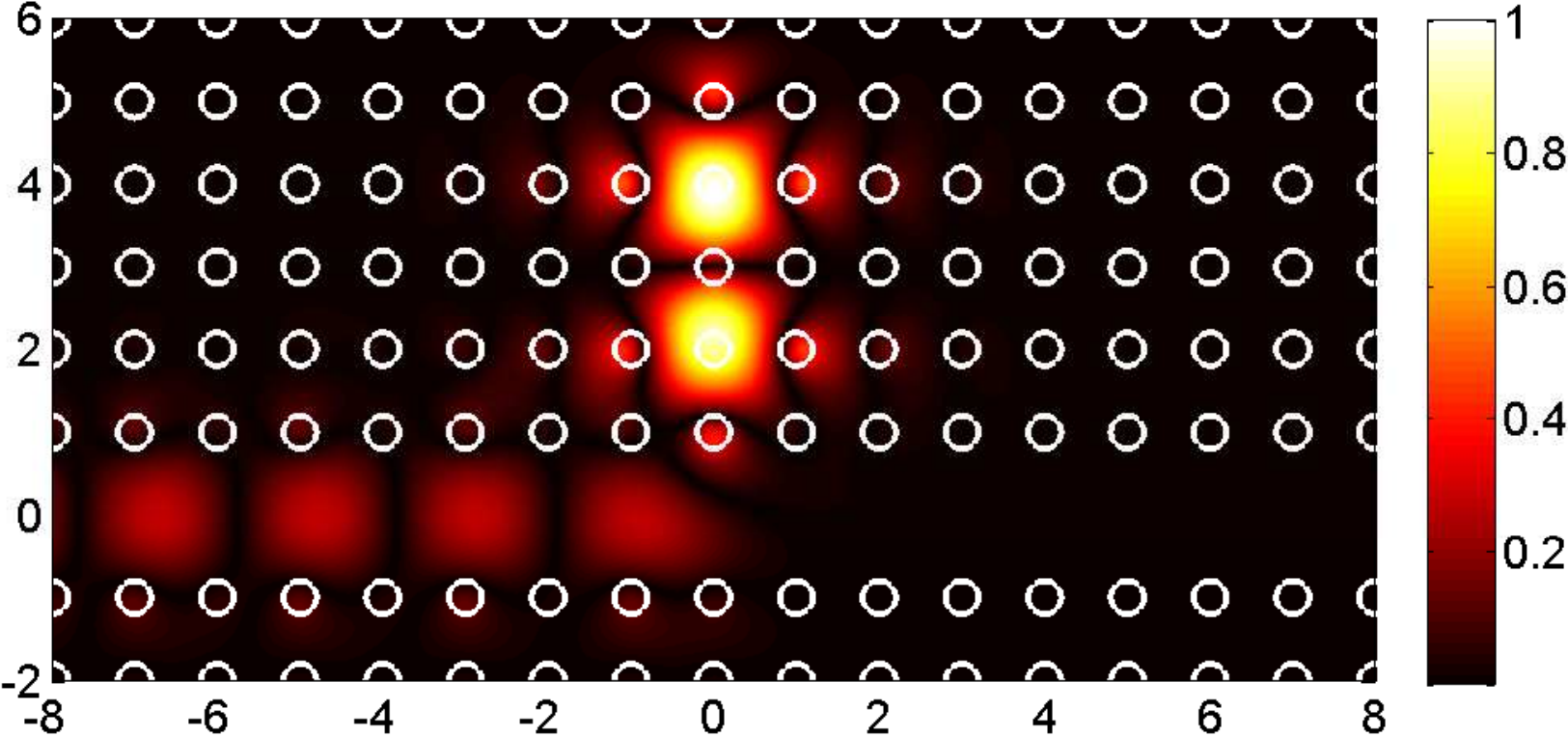}
\put(35,-7){Position $x/a$}
\put(-8,14.5){\begin{sideways}{Position $y/a$}\end{sideways}}
\end{overpic}\\[9mm]
%
\begin{overpic}[width=7.37cm]{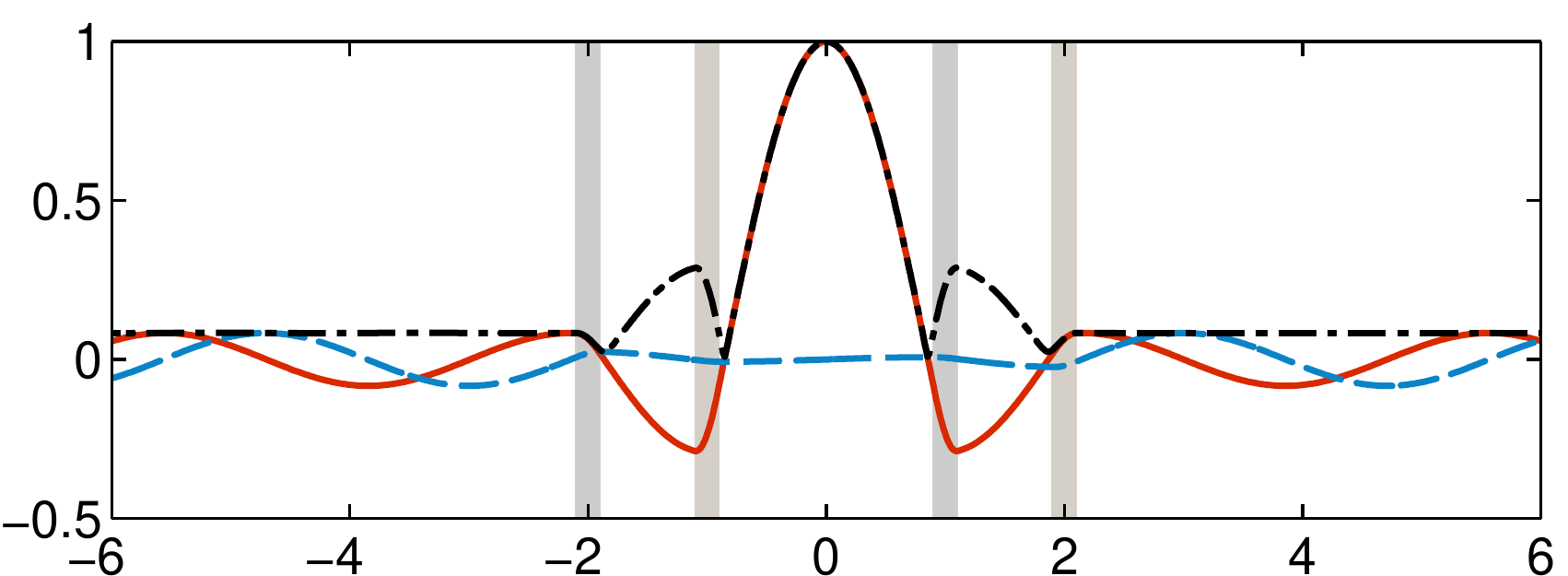}
\put(41.5,-7){Position $x/a$}
\put(-5,6){\begin{sideways}{Relative field}\end{sideways}}
\end{overpic}\quad\quad\;\;\,\\[11mm]
\begin{overpic}[width=7.2cm]{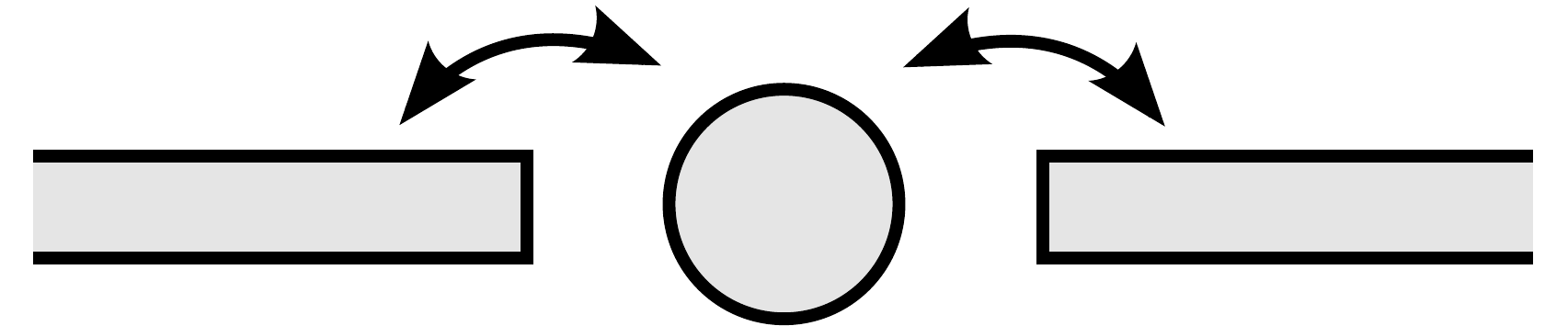}
\put(45.7,6.5){$A(t)$}
\put(3,14){$S_{1+}(t)\rightarrow$}
\put(3,-2){$\leftarrow S_{1-}(t)$}
\put(79,14){$\leftarrow S_{2+}(t)$}
\put(78,-2){$S_{2-}(t)\rightarrow$}
\put(27,21){$\gamma_\text{c}/2$}
\put(64,21){$\gamma_\text{c}/2$}
\end{overpic}\quad\quad\,\\[2mm]
\caption{\label{Fig:exampleFilter}Top: Example transmission calculation for a coupled cavity-waveguide structure in a two dimensional PC. The color coding indicates the relative strength of the out-of-plane electric field when illuminated from the left channel at the frequency $\omega a/2\pi\text{c}=0.385$. Center: One dimensional example of the transmission through a finite sized PC, showing the real (red solid) and imaginary (blue dashed) parts as well as the absolute value (black dashed-dotted) of the relative electric field $E(x)/E(0)$ when illuminated from the left at the cavity resonance frequency. Bottom: General schematic for use in setting up the CMT model for an in-line coupled cavity.\\[-4mm]
} 
\end{figure}
In general, an exhaustive description of the coupled cavity-waveguide dynamics requires the solution of a highly complicated set of  partial differential equations in time and space for the electromagnetic field as well as possible auxiliary equations for carrier transport, heat diffusion or other important processes. The dynamical equations governing each of the processes are well known, and such an approach is therefore in principle possible by numerical means. In practice, however, the complexity of the numerical model makes the computational requirements prohibitively large and limits the accuracy and the parameter ranges to be studied as well as physical insight to be gained. As an alternative approach to describe the fundamental dynamics in an effective and physically transparent way, coupled cavity-waveguide systems are often modeled by use of a rate equation model known as (temporal) coupled mode theory (CMT) ~\cite{Pierce_JAP_25_179_1954, Haus_Proc_IEEE_79_1505_1991, Haus_1984,Joannopoulos2008}. CMT aims to set up systems of ordinary differential equations to model light propagation in optical micro structures and circuits consisting of waveguides and cavities. This obviously represents an enormous simplification compared to the full set of Maxwell's equations. 
Nevertheless, the approach works remarkably well and provides an intuitive and physically appealing framework for the study and design of optical circuit elements and cavity based experiments~\cite{Soljacic_PRE_66_055601_2002, Yanik_OL_28_2506_2003,Soljacic_NatureMat_3_211_2004, Maes_JOSAB_22_1778_2005, Dumeige_OC_250_376_2005, Johnson_OE_14_817_2006, Bravo-Abad_JLT_25_2539_2007, deRossi_PRA_79_043818_2009, Kristensen_APL_102_041107_2013, Yu_OE_21_31047_2013, Heuck_OL_38_2466_2013, Mork_PRL_113_163901_2014,Yu_OE_40_2357_2015,Moille_PRA_94_023814_2016}. Moreover, the connection to quantum field theory was pointed out at an early stage~\cite{Lousell_JAP_33_2435_1962} and a framework similar to that of CMT forms the basis of several theoretical approaches to quantum optics with coupled cavity-waveguide systems~\cite{Gardiner_PRA_31_3761_1985, Ho_PRE_58_2965_1998, Cirac_PRL_78_3221_1997, Duan_PRL_92_127902_2004}.

Figure.~\ref{Fig:exampleFilter} shows examples of typical transmission type calculations for coupled cavity-waveguide structures in one and two dimensions as well as a typical schematic representation of the coupling between different parts of the system in the case of an in-line configuration such as in the one dimensional example. 
The standard approach of CMT is to set up equations coupling the temporal field amplitudes $A(t)$ and $S_{n\pm}(t)$. The fields are normalized so that $|A(t)|^2$ represents the optical energy inside the cavity, and $|S_{n\pm}(t)|^2$ represents the optical power in waveguide $n$ that is transmitted towards ($+$) or away from ($-$) the cavity. With these definitions, the CMT equations for the one dimensional example take the form~\cite{Joannopoulos2008}
\begin{align}
\frac{\ud}{\ud t}A(t) & = -\text{i}\omega_\text{c}(t) - \gamma_\text{c} A(t) + \sqrt{\gamma_\text{c}}\,S_{n+}(t) \label{Eq:linearFieldEq}\\
S_{n-}(t) & = S_{n+}(t)-\sqrt{\gamma_\text{c}}\,A(t) , \label{Eq:wglOutput}
\end{align}
where 
$\omega_\text{c}$ represents the cavity resonance frequency and $\gamma_\text{c}$ is the decay rate of the field in the cavity. This simple one cavity case may be readily generalized, and multimode cavities with multiple ports are discussed in Ref.~\cite{Suh_JQE_40_1511_2004}. In deriving Eqs.~(\ref{Eq:linearFieldEq}) and (\ref{Eq:wglOutput}), no details are given of the cavity and waveguide fields and no formal definition is given of the boundaries of the cavity. This reflects the enormous applicability of the equations, which were derived using only the assumptions of energy conservation and the existence of a resonance in a system with a driven harmonic oscillator~\cite{Haus_1984,Joannopoulos2008}. The simplicity, however, comes at a price when the theory is scaled to larger systems of coupled waveguides and cavities, \ptk{such as in the top panel of Fig.~\ref{Fig:exampleFilter}}, because the lack of definition of the cavity boundaries effectively means that the phases of the coupling coefficients are unknown. Although this is of limited concern in the single cavity case, it can 
be important in the modeling of coupled cavities or devices relying on interference based effects. 
Moreover, the incorporation of additional physical effects in the model 
typically relies on the electric field strength, and therefore requires additional scaling of the field amplitudes $A(t)$, which implicitly must depend on the cavity boundaries because of the choice of normalization. In addition to the practical limitations of Eqs.~(\ref{Eq:linearFieldEq}) and (\ref{Eq:wglOutput}), there is an inherent and intriguing mathematical difficulty in the fact that the CMT equations seem to couple the fields in a conservative system (the waveguide) with those in a non-conservative system (the cavity).

In this Article, we provide an alternative derivation of the CMT equations which originates from properties of the two families of modes that can be unambiguously defined for the waveguides and the cavities, respectively. As in Eqs.~(\ref{Eq:linearFieldEq}) and (\ref{Eq:wglOutput}), the end result 
does not rely on a definition of the cavity boundaries, 
but does include definite, and in general non-trivial, phase relations between the different modes. 
We discuss and illustrate how the cavity modes can be unambiguously defined as quasinormal modes (QNMs), which are solutions to a non-Hermitian eigenvalue problem and have complex resonance frequencies~\cite{Ching_1996,Kristensen_ACSPhot_1_2_2014}. 
The derivations of the CMT equations are based on the field equivalence principle, by which an incoming waveguide mode acts as a source for the QNM of the cavity. Similar ideas were introduced for transmission calculations in a one dimensional system in Ref.~\cite{Settimi_JOSAB_26_876_2009}, \ptk{and very recently a scattering matrix approach to the CMT was presented in Ref.~\cite{Alpeggiani_arXiv_1609_03902v1}.} 
The calculations result in generalized CMT equations for the (complex) electric field amplitudes which are similar in complexity to the standard CMT equations, and for the one dimensional example system in Fig.~\ref{Fig:exampleFilter} we show explicitly how the theory reduces to  Eqs.~(\ref{Eq:linearFieldEq}) and (\ref{Eq:wglOutput}) in the case of cavities with high $Q$-values.

The Article is organized as follows: In section~\ref{Sec:Theory} we set up the theoretical framework and derive the CMT equations. We first define the modes of the waveguides and cavities and discuss their normalization and use in expansion of the electric field Green tensor. We then use the field equivalence principle to formulate the coupling between the two types of modes; this forms the basis for a derivation of the CMT equations, in which the waveguide modes act as a driving term for the cavity modes as in Eq. (\ref{Eq:linearFieldEq}). 
Section~\ref{Sec:Examples} provides example calculations in one and two dimensions, where we assess the validity of the theory by comparing directly to high accuracy reference calculations. Last, we present the conclusions in Section~\ref{Sec:Conclusion}.


\section{Theory}
\label{Sec:Theory}
In this section we derive the 
CMT equations. First, we introduce rigorous definitions of the modes in both the waveguides and the cavity. Next, we calculate the coupling between the modes in the waveguides and the cavities and derive the CMT equations. 
\hmm{Last, we use the theory of non-Hermitian perturbation theory to include the effects of small changes in the permittivity - but this is well known, so probably not.}

\subsection{Definition of modes}
\label{sec:Modes}
In general, we define the different modes of the subsystems to be time-harmonic solutions to the source-free Maxwell equations of the form
\begin{align}
\mf(\mr,t) = \mf(\mr,\omega)\text{e}^{-\text{i}\omega t},
\end{align}
where the position dependent field $\mf(\mr,\omega)$ solves the wave equation
\begin{align}
\nabla\times\nabla\times \mf(\mr,\omega)-k^2\epsilon_\text{r}(\mr)\mf(\mr,\omega) = 0,
\label{Eq:HelmholtzEq}
\end{align}
in which $\epsilon_\text{r}(\mr, \omega)$ is the position dependent relative permittivity and $k=\omega/\text{c}$ is the ratio of the angular frequency to the speed of light. For simplicity, we limit the analysis to non-dispersive materials. The wave equation alone does not suffice to unambiguously define the modes --- only by specifying a suitable set of boundary conditions do we get a differential equation problem with corresponding solutions that we might define as modes. The proper choice of boundary conditions depends on the specific subsystem, and 
we argue that different choices are appropriate for the waveguide modes and the cavity modes. 

\subsubsection{Waveguide modes}
We consider general waveguides 
for which 
the relative permittivity can be written as $\epsilon_\text{r}(\mr+\mathbf{R}) = \epsilon_\text{r}(\mr)$, where $\mathbf{R}$ is a lattice vector in the direction of the waveguide. The proper boundary conditions in this case are periodic boundary conditions
with an optional phase. 
Bloch-Floquet theory ensures that the solutions may be written as $\mathbf{F}_\mk(\mr)=\mf_\mk(\mr)/\sqrt{L}$, where $L$ is the length of the normalization volume, and
\begin{align}
\mf_\mk(\mr) = \text{e}^{\text{i}\mk\cdot\mr}\mathbf{u}_\mk(\mr),
\label{Eq:Blochform}
\end{align}
in which $\mathbf{u}_\mk(\mr+\mathbf{R}) = \mathbf{u}_\mk(\mr)$ \jesm{is the Bloch function and $\mk$ is the wave vector in the direction of the waveguide}. \Jakob{As a special case, Eq.~(\ref{Eq:Blochform}) applies also to translationally invariant waveguides, such as optical fibers for example, for which $\mathbf{u}_\mk(\mr)$ is independent of the position along the waveguide. Using periodic boundary conditions,} 
Eq.~(\ref{Eq:HelmholtzEq}) yields a Hermitian differential equation problem for the Bloch functions $\mathbf{u}_\mk(\mr)$ \Jakob{in a single waveguide unit cell}, which may be solved analytically in certain cases or by standard numerical methods such as plane wave expansion~\cite{Johnson_OE_8_173_2001} or 
finite elements (FEM)~\cite{Iserles_2003}. 
The full solution to a waveguide geometry problem comprises both guided modes, 
which decay exponentially in the direction perpendicular to the waveguide, and radiation modes, 
which oscillate in the direction perpendicular to the waveguide and may or may not decay in the propagation direction~\cite{SnyderAndLove_2000}. For the present purpose, however, we consider only the guided modes\jesm{, for which the wave vector $\mk$ is real}, 
and we limit the analysis to waveguides that support only a single band of guided modes with an approximate linear dispersion in the bandwidth of interest. 
The guided modes are normalized as
\begin{align}
\int_V\epsilon_\text{r}(\mr)\mathbf{F}_\mk^*(\mr)\cdot\mathbf{F}_\mathbf{q}(\mr)\,\ud V = \delta_{\mk\mathbf{q}},
\label{Eq:normalization_of_f}
\end{align}
where $V=L^3$ is \Jakob{the normalization volume} and $\delta_{\mk\mathbf{q}}$ is the Kronecker delta function. From Eqs.~(\ref{Eq:Blochform}) and (\ref{Eq:normalization_of_f}) we then find that
\begin{align}
\frac{1}{a}\int_\text{UC}\epsilon_\text{r}(\mr)\mathbf{u}^*_\mk(\mr)\cdot\mathbf{u}_\mathbf{q}(\mr)\,\ud V =\delta_{\mk\mathbf{q}},
\label{Eq:blochModeNormalization}
\end{align}
where the integral is over the volume of a single waveguide unit cell of length $a$. 
In general, we may expand any time-dependent electric field, which is guided in the waveguide $n$, as a sum over the guided modes as
\begin{align}
\mE_{n\pm}(\mr,t) &= 
E_n\sum_\omega \zeta(x,\omega)e^{-\text{i}\omega t}\mf_{n\pm}(\mr,\omega) ,
\label{Eq:Field_CMT_form_WG}
\end{align}
where $\pm$ denotes the direction of propagation. 
The coordinate $x$ specifies the position along the waveguide, \jesm{so that $\zeta(x,\omega)$ governs the modulation of the mode functions $\mf_{n\pm}(\mr,\omega)$ throughout the waveguide; the field variation transverse to the waveguide} 
is entirely contained in the mode functions. 
For any waveguide mode $\mf_{n+}(\mr,\omega)$, we fix the phase of the waveguide mode traveling in the opposite direction to be $\mf_{n-}(\mr,\omega)=\mf_{n+}^*(\mr,\omega)$. With this phase convention, the guided mode contribution to the Green tensor in the waveguide $n$ may be written as
\begin{align}
\mG_\text{wg}(\mr,\mr',\omega) &\approx \frac{\text{\text{i}}}{2k}\frac{\text{c}}{v_\text{g}}\Big[ \Theta(x-x')\mf_{n+}(\mr,\omega)\mf_{n+}^*(\mr',\omega) \nonumber \\
 &\;\;\qquad+ \Theta(x'-x)\mf_{n-}(\mr,\omega)\mf_{n-}^*(\mr',\omega) \Big],
\label{Eq:GreensTensorWG}
\end{align}
where \jesm{$k=|\mk|$,} $v_\text{g}=\partial\omega/\partial k$ is the group velocity and $\Theta(x)$ is the Heaviside step function. 



\subsubsection{Cavity modes}
\label{Sec:cavityModes}
Optical cavities are fundamentally different from waveguides because they act as resonators for light at discrete frequencies, and because they are inherently leaky~\cite{Ching_1996,Kristensen_ACSPhot_1_2_2014} resulting in an exponential decay of energy in the cavity over time. 
 The leaky nature of the cavity modes can be conveniently modeled by use of a 
radiation condition in the defining differential equation problem. Augmenting Eq.~(\ref{Eq:HelmholtzEq}) with a radiation condition leads to a non-Hermitian eigenvalue problem, and the solutions are QNMs~\cite{Ching_1996,Kristensen_ACSPhot_1_2_2014} with discrete and complex resonance frequencies $\tlo_\mu=\omega_\mu-\text{i}\gamma_\mu$. For any surface $A$ with normal vector $\mathbf{n}$ fully enclosing an optical cavity with a single QNM $\mu$, the time-averaged power leaking through the surface, as given in terms of the Poynting vector $\mathbf{S}(\mr)$, is related to the integral of the time-averaged energy density $u(\mr)$ inside the surface as 
\begin{align}
\int_A\langle\mathbf{S}(\mr)\rangle\cdot\mathbf{n}\,\ud A = 2\gamma_\mu\int_V \langle u(\mr)\rangle\,\ud V,
\label{Eq:power_energy_ratio}
\end{align}
from which the $Q$-value may be written as $Q=\omega_\mu/2\gamma_\mu$~\cite{Lalanne_LaserPhotRev_2_514_2008}. In geometries with a homogeneous permittivity distribution $\epsilon_\text{B}=n_\text{B}^2$ at large distances \Jakob{from the cavity}, the proper choice of radiation condition is \ptk{arguably} the Silver-M{\"u}ller radiation condition~\cite{Kristensen_ACSPhot_1_2_2014, Martin_MultipleScattering} in the form~\cite{Kristensen_PRA_92_053810_2015}
\begin{align}
\hat{\mr}\times\nabla\times\mft_\mu(\mr) \rightarrow -\text{i}n_\text{B}\tilde{k}_\mu\mft_\mu(\mr)\quad\text{as}\;r\rightarrow\infty,
\label{Eq:SilverMuller}
\end{align}
where $\hat{\mr}$ is a unit vector in the direction of $\mr$. 
For general coupled cavity-waveguide structures, however, 
the coupling to the waveguides often represents the largest decay channel for the optical energy in the cavity, and one cannot hope to accurately calculate the cavity mode without including the coupling to the waveguide. In the case of cavities coupled to periodic waveguides in PC membranes,
for example, one would require Eq.~(\ref{Eq:SilverMuller}) to be satisfied at positions above and below the membrane. For the \jesm{part of the field leaking through the} waveguides, one can enforce a waveguide radiation condition by demanding that the QNMs satisfy a condition similar to Eq.~(\ref{Eq:Blochform}), \jesm{but with the wave vector in each of the waveguides pointing away from the} 
cavity~\cite{Li_JOSA_B_26_2427_2009,deLasson_JOSA_A_31_2142_2014, Kristensen_OL_39_6359_2014}. \ptk{In particular, for positions in or near the waveguide, but sufficiently far away from the cavity that the influence of non-propagating waveguide modes can be neglected, the QNM can be written in terms of the analytical continuation of the waveguide mode traveling away from the cavity as
\begin{align}
\mft_\mu(\mr) = \sigma_{\mu n}\mf_{n-}(\mr,\tlo_\mu),
\label{Eq:sigmaTildeDef} 
\end{align}
where $\sigma_{\mu n}$ is a complex constant which depends on the choice of phase of the waveguide modes.} \ngre{See Appendix~\ref{App:Sigma_def} for details on the expansion of cavity modes in terms of waveguide modes.}

The use of radiation conditions ensures that light propagates away from the cavity as expected for a leaky resonator, but this comes at the price of a conceptually challenging property of the QNMs, namely the fact that they diverge (exponentially) in the limit $r\rightarrow\infty$. This exponential divergence in the spatial domain is a natural consequence of the propagating nature of electromagnetic fields in combination with the exponential temporal decay of the field in the cavity~\cite{Tikhodeev_PRB_66_045102_2002,deLasson_PhD_2015}. 
In practice, it has the important consequence that the QNMs cannot be normalized by the integral formula in Eq.~(\ref{Eq:normalization_of_f}) which is commonly adopted for Hermitian eigenvalue problems. For resonators in homogeneous backgrounds, the proper normalization has been derived in at least three different ways~\cite{Lai_PRA_41_5187_1990, Muljarov_EPL_92_50010_2010, Sauvan_PRL_110_237401_2013} which are closely related and provide the same result~\cite{Kristensen_PRA_92_053810_2015}. The differences in the normalization integrals can be understood as arising from different regularizations of an inherently ill-defined integral~\cite{Kristensen_PRA_92_053810_2015}. For cavities coupled to periodic waveguides, \Jakob{this observation} was used in Ref.~\cite{Kristensen_OL_39_6359_2014} as a \Jakob{motivation} for regularizing the normalization integral by means of the theory of divergent series. With such an approach, but writing the integral in terms of the electric field QNMs only, the QNMs may be normalized \comm{via the integral
\begin{align}
I&= \int_V \epsilon_\text{r}(\mr)\mft_\mu(\mr)\cdot\mft_\mu(\mr)\,\ud V,
\label{Eq:QNM_Norm}
\end{align}%
where the volume $V$ extends over all space, but is split into different parts; one part containing the cavity and one part for each of the infinite waveguides leading away from the cavity. Considering, for simplicity, a single waveguide extending from the cavity in the positive $x$ direction, 
the integral is split as $I=I_{x<x_0}+I_{x>x_0}$, and calculation of $I_{x<x_0}$ is performed over a volume extending to a distance along the waveguide, $x_0$, which is chosen sufficiently large that the QNM is well described by Eq.~(\ref{Eq:Blochform}) with a single complex wave vector $\tilde{\mk}_\mu$. For the region $x>x_0$, where the integrand diverges exponentially with increasing $x$, one can rewrite the integral as
\begin{align}
I_{x>x_0}&= I_a(x_0)\sum_{m=0}^\infty b^m,
\label{Eq:divergentGeometric}
\end{align}
where $b = \exp\{2\text{i}\tilde{k}_\mu a\}$ and $I_a(x_0)$ is an integral as in Eq.~(\ref{Eq:QNM_Norm}) but limited to a single unit cell along the waveguide direction from $x=x_0$ to $x=x_0+a$. Since $|b|>0$, the series in Eq.~(\ref{Eq:divergentGeometric}) is formally divergent. Nevertheless, one can use the theory of divergent series~\cite{Hardy_1949} to assign to it the finite value
\begin{align}
I_{x>x_0} = \frac{I_a(x_0)}{1-\text{e}^{2\text{i}\tilde{k}_\mu a}}.
\label{Eq:I_Lambda}
\end{align}
This} regularization procedure has been used for QNM perturbation calculations~\cite{Kristensen_OL_39_6359_2014} as well as QNM approximations to the local density-of-states~\cite{deLasson_OL_40_5790_2015,Malhotra_OE_24_13574_2016} in coupled cavity-waveguide systems. 




\hmm{For a spherically symmetric system, such as an optical micro sphere, Lee \emph{et al.} have shown that the QNMs provide a complete set of transverse modes~\cite{Lee_JOSAB_16_1409_1999}.  }


In many practical applications of QNMs, and in the present case in particular, we are not interested in a full expansion of the field. Rather, we seek an expansion in terms of at most a few QNMs in each cavity, which can be treated analytically and often provide a surprisingly accurate description~\cite{Settimi_JOSAB_26_876_2009, Kristensen_OL_37_1649_2012, Sauvan_PRL_110_237401_2013, Bai_OE_21_27371_2013, Kristensen_ACSPhot_1_2_2014, Ge_OL_39_4235_2014, Ge_NJP_16_113048_2014, deLasson_OL_40_5790_2015, Malhotra_OE_24_13574_2016, Sauvan_PRA_89_043825_2014}. In such an approach, at frequencies close to the cavity resonance and positions in or near the cavity, we assume that the Green tensor may be \ptk{well approximated as \cite{Sauvan_PRA_89_043825_2014}
\begin{align}
\mG_\text{cav}(\mr,\mr',k) \approx \sum_\mu\frac{\mft_\mu(\mr)\mft_\mu(\mr')}{2k (\tilde{k}_\mu-k)},
\label{Eq:GreensTensorCavity}
\end{align}
where $\mu$ runs over the (few) QNMs of interest for the given cavity. The denominator in Eq.~(\ref{Eq:GreensTensorCavity}) is slightly different from that of another expansion of the Green tensor in terms of the QNMs, where the denominator is replaced by $2\tilde{k}_\mu(\tilde{k}_\mu-k)$~\cite{Lee_JOSAB_16_1409_1999}. It was pointed out in Ref.~\cite{Muljarov_EPL_92_50010_2010}, that the two forms can be related by use of a 
QNM completeness relation. For the present purpose with an explicit truncation of the summation, we generally do not expect one to be more precise than the other, but the form in Eq.~(\ref{Eq:GreensTensorCavity}) results in slightly simpler expressions when used to derive the CMT equations below.} 
Finally, as in the case of the Green tensor, we also assume that one can approximate the time dependent field in the cavity in terms of the few QNMs of interest as
\begin{align}
\mE_{\text{cav}}(\mr,t) = \sum_\mu E_\mu(t)\mft_\mu(\mr).
\label{Eq:Field_CMT_form_Cavity}
\end{align}


\subsection{Derivation of the CMT equations}
Given that the waveguide modes and the cavity modes derive from differential equation problems with very different boundary conditions, it is not obvious how to calculate a coupling between them directly from Maxwell's equations. \Jakob{In particular, the cavity modes contain only outwards propagating field components, so it is unclear how one can calculate the coupling between the incoming light in the waveguide and the field in the cavity by use of an overlap integral, even though this is a well established approach for calculating the coupling between co-propagating beams of light in parallel waveguides. }
\Jakob{Clearly}, in deriving the CMT equations from energy conservation arguments~\cite{Joannopoulos2008} one elegantly avoids this 
challenge, \jesm{but at the price of leaving the phase of the coupling constant unspecified. It would therefore} 
be both useful and enlightening to have a formal understanding of how to connect the two types of modes. Below, we take an alternative approach and derive the coupling coefficient as well as the CMT equations by way of the two different approximations of the Green tensor in Eqs.~(\ref{Eq:GreensTensorWG}) and (\ref{Eq:GreensTensorCavity}).




\subsubsection{Coupling waveguide modes with cavity modes}
\label{Sec:couplingFormalism}
We consider the case of an electromagnetic field, at a single frequency $\omega$ and incident in waveguide $n$, of the form
\begin{align}
\mE_\text{in}(\mr,\omega) &= E_{n+}(\omega)\mf_{n+}(\mr,\omega) \label{Eq:inputWGfield} \\
\mB_\text{in}(\mr,\omega) &= -\frac{\text{i}}{\omega}\nabla\times\mE_\text{in}(\mr,\omega),
\label{Eq:inputWGfield_B}
\end{align}
where the field $\mf_{n+}(\mr,\omega)$ has the Bloch form in Eq.~(\ref{Eq:Blochform}). The field carries the average input power
\begin{align}
P_{n+} &= \frac{1}{2\mu_0}\int_{D_n} \text{Re}\left\{\mE_\text{in}(\mr,\omega)\times\mB_\text{in}^*(\mr,\omega)\right\}\cdot\mathbf{n}\,\ud A,
\label{Eq:P1}
\end{align}
where $\mu_0$ is the permeability of free space, $D_n$ is any plane intersecting the waveguide and $\mathbf{n}$ is a unit vector in the direction of propagation. The input power equals the product of the group velocity and the unit cell average of the energy density~\cite{Joannopoulos2008}, so by Eqs. (\ref{Eq:blochModeNormalization}) and (\ref{Eq:inputWGfield})-(\ref{Eq:inputWGfield_B}) we may write it as
\begin{align}
P_{n+} &= \frac{v_\text{g}}{2L_\text{UC}}\int_\text{UC}\epsilon_0\epsilon_\text{r}(\mr,\omega)|\mE_\text{in}(\mr,\omega)|^2\,\ud V, \nonumber \\
&= \frac{v_\text{g}}{2}\epsilon_0 |E_{n+}(\omega)|^2,
\label{Eq:P1plus}
\end{align}
in which $\epsilon_0$ is the permittivity of free space.

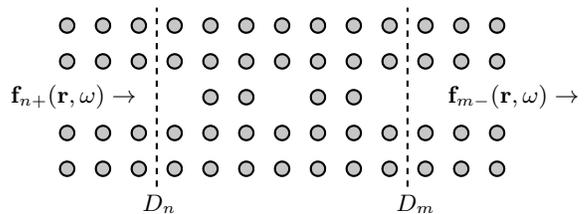
\begin{figure}[htb]
\centering
\begin{overpic}[width=6.25cm]{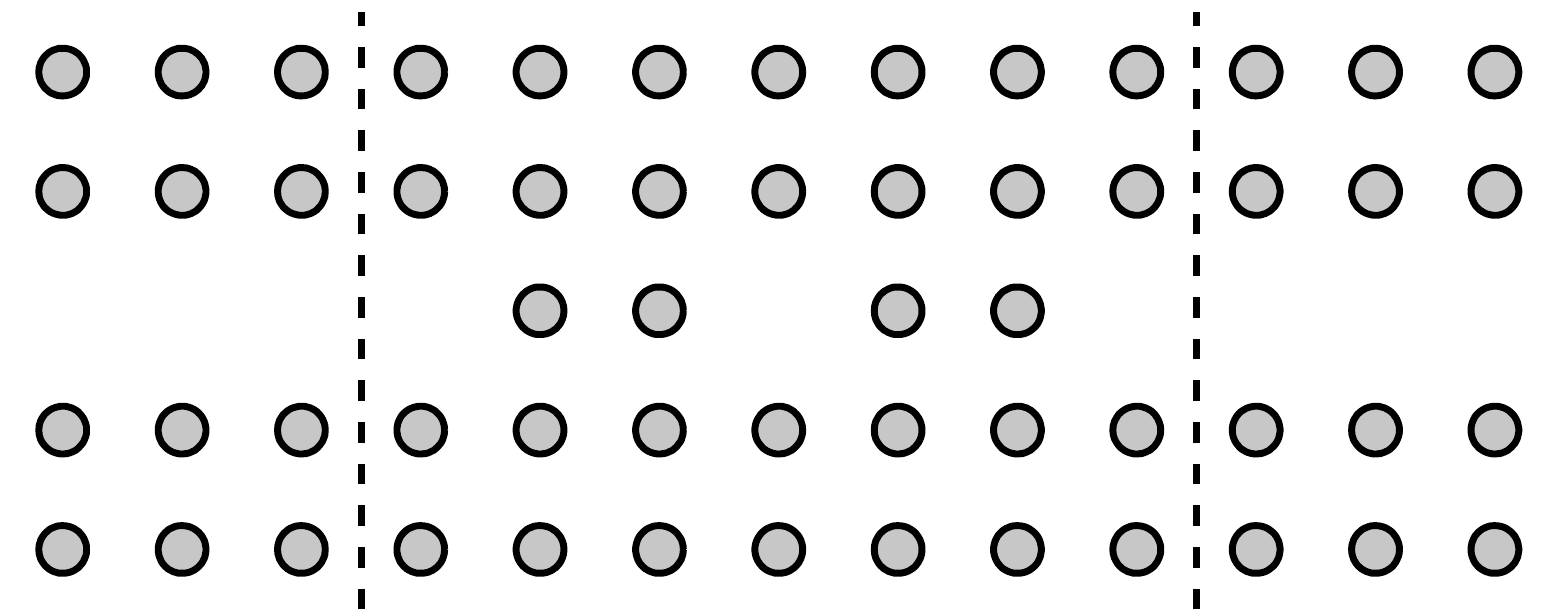}
\put(-8,18){$\mf_{n+}(\mr,\omega)\rightarrow$}
\put(85,18){$\mf_{m-}(\mr,\omega)\rightarrow$}
\put(20,-5){$D_n$}
\put(74,-5){$D_m$}
\end{overpic}\\[2mm]
\caption{\label{Fig:inlineCavity_geom}Cavity coupled to waveguides in an in-line configuration. }
\end{figure}

By use of the field equivalence principle~\cite{Jin_2010}, the incident field is identical to the field from electric and magnetic current sources in the plane $D$ of the form
\begin{align}
\mathbf{J}_\text{in}(\mr,\omega) &= \hat{\mathbf{n}}\times\left( \frac{1}{\mu_0}\mB_\text{in}(\mr,\omega)\right) \\
\mathbf{M}_\text{in}(\mr,\omega) &= \mE_\text{in}\times\hat{\mathbf{n}}.
\label{Eq:sourceCurrent}
\end{align}
Therefore, at any $x>x'$, we can write the electric field as~\cite{Jin_2010}
\begin{align}
\mE(\mr,\omega) &= \int_D\text{i}\omega\mu_0\mG(\mr,\mr',\omega)\cdot\mathbf{J}_\text{in}(\mr',\omega)\,\ud A\nonumber\\
&\quad - \int_D \big[\nabla\times\mG(\mr,\mr',\omega)\big]\cdot\mathbf{M}_\text{in}(\mr',\omega)\,\ud A
\label{Eq:EfromSourceCurrent}
\end{align}
where $\mG(\mr,\mr',\omega)$ is the electric field Green tensor for the particular geometry of interest.
As an example, for the infinite waveguide 
we can use Eq.~(\ref{Eq:EfromSourceCurrent}) with the waveguide Green tensor in Eq.~(\ref{Eq:GreensTensorWG}) to find that
\begin{align}
\mE_\text{wg}(\mr,\omega) = \frac{2\mu_0\text{c}^2P_{n+}}{v_\text{g}|E_{1+}(\omega)|^2}\mE_\text{in}(\mr,\omega) = \mE_\text{in}(\mr,\omega),
\label{Eq:Ewg_from_Equivalence}
\end{align}
confirming that the field at any point in the infinite waveguide is simply the input field.


In the case of a waveguide coupled to a cavity, as shown schematically in Fig.~\ref{Fig:inlineCavity_geom}, we can use the QNM expansion of the Green tensor in Eq.~(\ref{Eq:GreensTensorCavity}) to calculate the field in the cavity due to a source current in the plane $D_n$. 
To this end, we note that for positions in the waveguide, the functional form of the cavity mode in the waveguide is similar to the analytical continuation of the waveguide mode traveling in the direction away from the cavity, \ptk{cf. Eq.~(\ref{Eq:sigmaTildeDef}). To a first approximation, and at positions in waveguide $n$ close to the cavity,} \ngre{we may neglect the change in the waveguide mode profile due to the relatively small imaginary part of the complex resonance frequency and set}
\begin{align}
\mft_\mu(\mr) \approx \sigma_{\mu n}\mf_{n-}(\mr,\omega_\mu),
\label{Eq:sigmaDef}
\end{align}
which is the key relation underlying most of the calculations below. This approximation is expected to be best at positions close to the cavity, as 
discussed in Appendix~\ref{Sec:CouplingParameter}. %
\ptk{With Eq.~(\ref{Eq:sigmaDef}), we can use the QNM expansion of the Green tensor in Eq.~(\ref{Eq:GreensTensorCavity}) and the field equivalence principle in Eq.~(\ref{Eq:EfromSourceCurrent}) to express the cavity field in terms of the input field as
\begin{align}
\mE_\text{cav}(\mr,\omega) &= \sum_\mu \text{i}\frac{v_\text{g}}{\omega-\tlo_\mu}\sigma_{\mu n} E_{n+}(\omega)\mft_\mu(\mr).
\label{Eq:cavityField}
\end{align}
}

%

We define the complex coupling of a general input field $\mE_\text{in}(\mr,\omega)=E_{n+}(\omega)\mf_{n+}(\mr,\omega)$ to the cavity field of a single QNM $\mE_\mu(\mr,\omega)=E_\mu(\omega)\mft_\mu(\mr)$ as 
\begin{align}
C_{\mu n}(\omega) = \frac{E_\mu(\omega)}{E_{n+}(\omega)}.
\end{align}
With this definition, we can express the amplitude of the cavity mode $\mu$ due to coupling of the input field in Eq.~(\ref{Eq:inputWGfield}) as
\begin{align}
E_\mu(\omega) = C_{\mu n}(\omega)E_{n+}(\omega),
\label{Eq:E_cav_from_E_wg}
\end{align}
where
\begin{align}
C_{\mu n}(\omega) = \text{i}\frac{v_\text{g}}{\omega-\tlo_\mu}\sigma_{\mu n} = \Gamma_\mu(\omega)\sigma_{\mu n}.
\label{Eq:C_def}
\end{align}


\subsubsection{Transmission and reflection from cavities}
In many practical situations, we are interested in the transmission and reflection from the cavity sections. In these situations, we must find a suitable approximation to the Green tensor for use in the field equivalence principle via 
Eq.~(\ref{Eq:EfromSourceCurrent}). The most general linear scattering of an incoming signal includes both coupling into the cavity and scattering into an outgoing signal through the same channel. By such arguments, additional reflected light is automatically built into the CMT~\cite{Joannopoulos2008} as can be seen directly from Eq. (\ref{Eq:wglOutput}). 
In keeping with the spirit of the current attempt at deriving the CMT directly from the properties of the modes, we include additional scattering and transmission contributions to the Green tensor 
by means of the Dyson equation, as detailed in Appendix~\ref{App:Dyson}. With such an approach, 
we can derive a general approximation of the Green tensor in the physically appealing form
\begin{align}
\mG(\mr,\mr',\omega) \approx \mG_\text{B}(\mr,\mr',\omega) + \sum_\mu\frac{\mft_\mu(\mr)\mft_\mu(\mr')}{2k(\tlk_\mu-k)},
\label{Eq:G_overall_approx}
\end{align}
as the sum of the Green tensor of the reference structure with no cavity, $\mG_\text{B}(\mr,\mr',\omega)$, and the QNM approximation of the cavity Green tensor.

For a general input field $\mE_\text{in}(\mr,\omega)=E_{n+}(\omega)\mf_{n+}(\mr,\omega)$ we define the complex transmission to the output channel $\mE_{m-}(\mr,\omega)=E_{m-}(\omega)\mf_{m-}(\mr,\omega)$ as 
\begin{align}
T_{mn}(\omega) = \frac{E_{m-}(\omega)}{E_{n+}(\omega)}.
\label{Eq:Trans_def}
\end{align}
For the transmission through an in-line coupled single-mode cavity, as in Fig.~\ref{Fig:inlineCavity_geom}, we can neglect the contribution from the background Green tensor $\mG_\text{B}(\mr,\mr',\omega)$ in Eq.~(\ref{Eq:G_overall_approx}) and express the transmission as
\begin{align}
T_{mn}(\omega)   = \text{i}\frac{v_\text{g}}{\omega-\tlo_\mu}\sigma_{\mu n}\sigma_{\mu m}= \Gamma_\mu(\omega)\sigma_{\mu n}\sigma_{\mu m}.
\label{Eq:T_inline_singleMode}
\end{align}
Alternatively, we can calculate the transmission through the in-line coupled single-mode cavity by use of the waveguide Green tensor and considering the cavity mode as the source term for fields in the output waveguide. \ptk{This gives the same result}.

As a straightforward generalization of the transmission, 
and keeping in mind the phase convention for the waveguide modes, we define the complex reflection coefficient to the output channel $\mE_\text{out}(\mr,\omega)=E_{n-}(\omega)\mf_{n+}^*(\mr,\omega)$ as
\begin{align}
R_n(\omega) = T_{nn}(\omega) = \frac{E_{n-}(\omega)}{E_{n+}(\omega)}.
\label{Eq:Reflect_def}
\end{align}
In the case of the in-line coupled single-mode cavity, we can write the approximate form for the Green tensor in Eq.~(\ref{Eq:G_overall_approx}) in the form
\begin{align}
\mG(\mr,\mr',\omega) \approx \frac{\text{i}}{2k}\frac{\text{c}}{v_\text{g}}\mf_{n-}(\mr)\mf_{n+}^*(\mr')\Phi + \frac{\mft_\mu(\mr)\mft_\mu(\mr')}{2k(\tlk_\mu-k)},
\label{Eq:G_scat_inline_cmt}
\end{align}
where $\Phi$ denotes a general phase which depends on the particular geometry and the phase of the waveguide modes. In the case of lossless reflection from perfectly conducting plates and the choice of phase where the waveguide modes are entirely real at the plates, the phase must be $\Phi=-1$. Using Eq.~(\ref{Eq:G_scat_inline_cmt}), we can follow calculations analogous to those for the straight waveguide in Eq.~(\ref{Eq:Ewg_from_Equivalence}) and the transmission through the single mode cavity in Eq.~(\ref{Eq:T_inline_singleMode}) to express the reflection as
\begin{align}
R(\omega)  = \Phi + \text{i}\frac{v_\text{g}}{\omega-\tlo_\mu}\sigma_{\mu n}^2= \Phi + \Gamma_\mu(\omega)\sigma_{\mu n}^2.
\label{Eq:R_inline_singleMode}
\end{align}


\subsubsection{The CMT equations}
The CMT equations follow immediately from Eq.~(\ref{Eq:cavityField}) by transformation to the time domain. For each term in the sum, we can Fourier transform to find
\begin{align}
E_\mu(t) &= \frac{1}{2\pi}\int_{-\infty}^\infty\text{i}\frac{v_\text{g}}{\omega-\tlo_\mu}\sigma_{\mu n} E_{n+}(\omega)\text{e}^{-\text{i}\omega t}\,\ud\omega.
\label{Eq:temporal_field_as_fourier_intergral}
\end{align}
Next, we multiply by $\exp\{i\tlo_\mu t\}$ and differentiate to find
\begin{align}
\frac{\ud}{\ud t}E_\mu(t) = -\text{i}\tlo_\mu E_\mu(t) + \text{i} v_\text{g} \sigma_{\mu n} E_{n+}(t).
\label{Eq:traditonal_CMT_with_sigma}
\end{align}
In order to connect to the standard formulation of CMT,  we 
consider a single cavity with high $Q$-value, for which
%
\hmm{
%
In addition, we drop the (small) last term in Eq.~(\ref{Eq:proper_CMT_with_correction}) to arrive at the simplified equation
\begin{align}
\frac{\ud}{\ud t}E_\text{cav}(t) = -\text{i}\tlo_\text{c}E_\text{cav}(t) + \sqrt{\gamma_\text{c}v_\text{g}}E_0\varphi(x_\text{s},t)e^{-\text{i}\omega_\text{L}t},
\label{Eq:traditonal_CMT_no_correction}
\end{align}
which 
corresponds to a single Lorentzian coupling as in Eq~(\ref{Eq:gamma_Lorentzian}). 
}
%
%
we can express the norm of $\sigma_{\mu n}$ in terms of physical parameters as $|\sigma_{\mu n}|^2\approx 2\gamma_n/v_\text{g}$, where $\gamma_n$ is the decay rate of the cavity mode though waveguide $n$; see Appendix~\ref{App:Energy} for details. In this case, for a single symmetric two-port cavity as in Fig.~\ref{Fig:inlineCavity_geom} with $\gamma_n=\gamma_\mu/2$, for example, we can choose the phases of the waveguide modes so that the coupling in Eq.~(\ref{Eq:C_def}) can be written as
\begin{align}
C_{\mu n}(\omega) &\approx 
\text{i}\frac{\sqrt{\gamma_\mu v_\text{g}}}{\omega-\tlo_\mu},
\label{Eq:gamma_Lorentzian}
\end{align}
which is exactly the typical Lorentzian coupling that one would expect from energy conservation arguments. For high-$Q$ cavities, the phase of the QNMs can be chosen so that the fields are almost entirely real. In this limit, the QNM norm in Eq.~(\ref{Eq:QNM_Norm}) reduces to the integral over the real energy density, so that we can  
use Eq.~(\ref{Eq:Field_CMT_form_Cavity}) to write the time-averaged energy in the cavity as
\begin{align}
U_\text{cav}(t) &\approx\frac{1}{2}\epsilon_0|E_\mu(t)|^2,
\end{align}
and since the QNMs are normalized, we can define
\begin{align}
A(t) = \sqrt{\frac{\epsilon_0}{2}}E_\mu(t),
\end{align}
so that $|A(t)|^2 = U_\text{cav}(t)$. Moreover, from Eq.~(\ref{Eq:P1plus}) we can immediately define
\begin{align}
S_{n+}(t) = \sqrt{\frac{\epsilon_0v_\text{g}}{2}}E_{n+}(t),
\end{align}
so that $|S_{n+}(t)|^2=P_{n+}(t)$. With these definitions, we can now multiply in Eq.~(\ref{Eq:traditonal_CMT_with_sigma}) by $\sqrt{\epsilon_0/2}$ to rewrite it in the exact form of Eq.~(\ref{Eq:linearFieldEq}). Similarly, to calculate the reflected light in the limit of high $Q$-values, we can set $\Phi=-1$ in Eq.~(\ref{Eq:R_inline_singleMode}) and use the expression for the coupling in Eq.~\ref{Eq:C_def} to write
\begin{align}
E_{n-}(\omega) = -E_{n+}(\omega) + E_\mu(\omega)\sigma_{\mu n}.
\end{align}
Multiplying by $\sqrt{\epsilon_0v_\text{g}/2}$ and transforming to the time domain, we can then define $S_{n-}(t)=-\sqrt{\epsilon_0v_\text{g}/2}E_{n-}(t)$ to recover the exact form of Eq.~(\ref{Eq:wglOutput}).

\section{Examples}
\label{Sec:Examples}
The CMT equations obviously provide an enormous simplification in practical calculations when compared to the full solution of the time-dependent Maxwell equations. The question remains as to the price, in terms of accuracy, one has to pay for this simplification. In this section we quantify the error by comparing CMT calculations to independent numerical solutions.
In Section~\ref{Sec:Example_1D}, we start by considering a one dimensional model system which 
captures most of the essential physics and has several advantages over higher-dimensional models: It is easy to clearly and exhaustively specify the model,  the QNMs are particularly easy to calculate, normalize and visualize and it is directly amenable to high-accuracy numerical verification by comparing to full calculations of Maxwell's equations. \comm{The cavity was previously used as an example for non-linear switching in Ref.~\cite{Kristensen_APL_102_041107_2013}, but without the explicit knowledge of the phase of the coupling as provided with the present theory}. As a second example, in Section~\ref{Sec:Example_2D}, we consider the two dimensional problem of a side-coupled cavity next to a PC waveguide, for which the properties of the QNM in the cavity was previously investigated in Refs.~\cite{Kristensen_OL_39_6359_2014} and \cite{deLasson_OL_40_5790_2015}. \ptk{To illustrate the usefulness of the theory in describing systems of coupled cavities, we also extend the example and 
calculate the transmission through the system with two side coupled cavities in the top panel in Fig.~\ref{Fig:exampleFilter}.}

\subsection{One dimensional example}
\label{Sec:Example_1D}
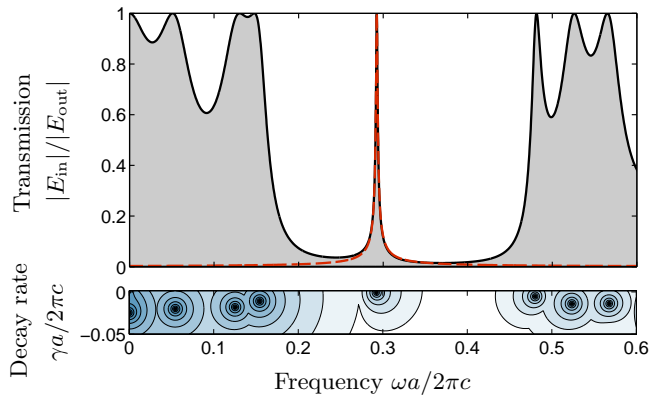
\begin{figure}[htb]
\flushright
\begin{overpic}[width=7.2cm]{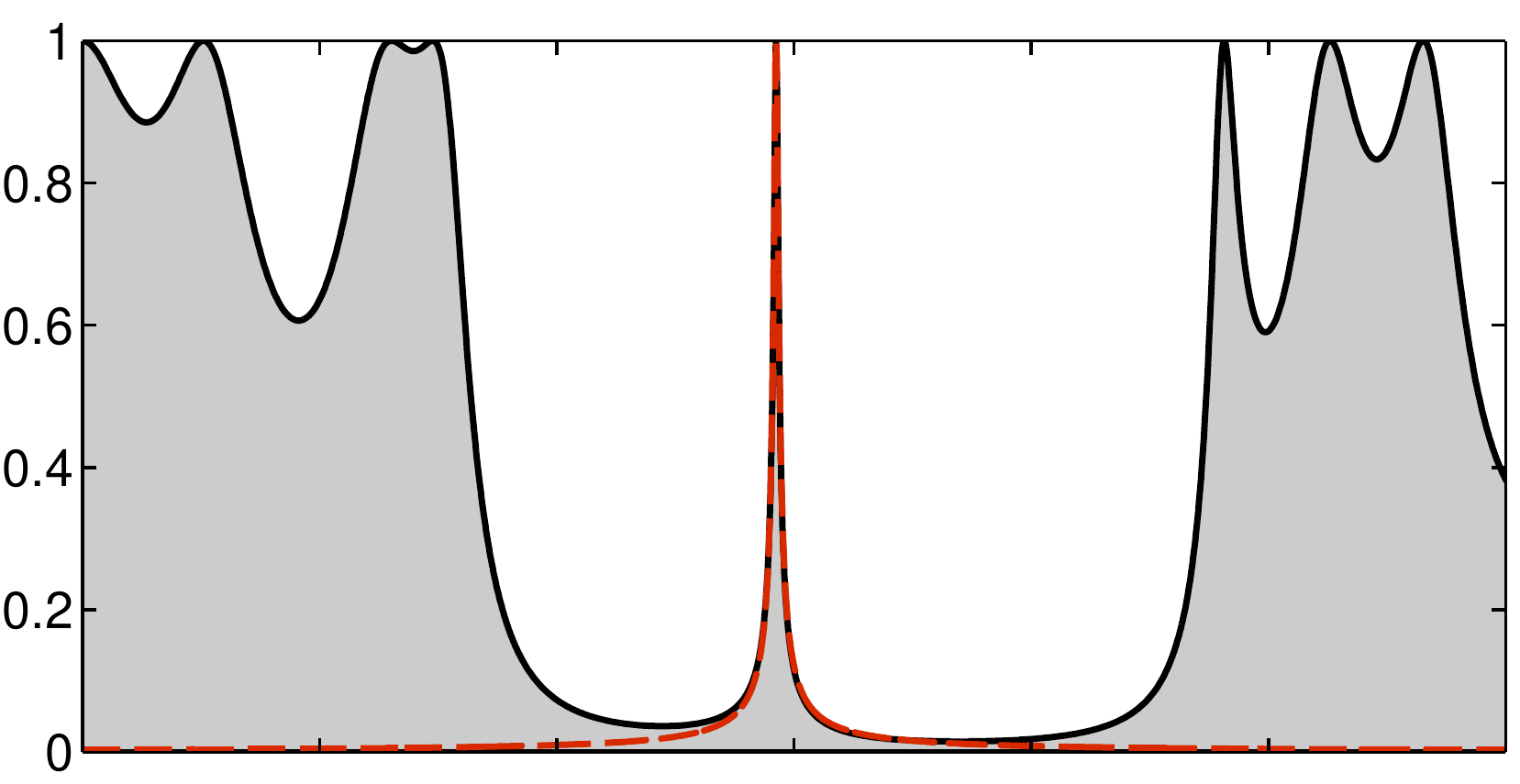}
\put(-16,12){\begin{sideways}{Transmission}\end{sideways}}
\put(-10,14){\begin{sideways}{$|E_\text{in}|/|E_\text{out}|$}\end{sideways}}
\end{overpic}\;\,\\
\begin{overpic}[width=7.62cm]{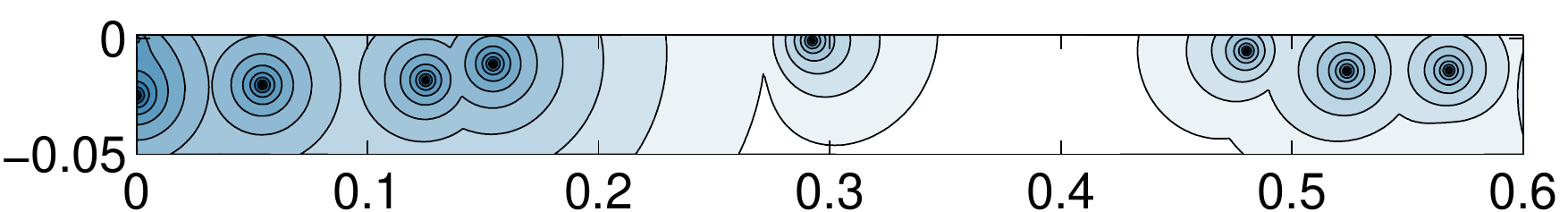}
\put(34,-6){Frequency $\omega a/2\pi c$}
\put(-12,-4){\begin{sideways}{Decay rate}\end{sideways}}
\put(-6,-1){\begin{sideways}{$\gamma a/2\pi c$}\end{sideways}}
\end{overpic}\\[6mm]
\caption{\label{Fig:cavity_1D_Q_160_pwTransmission}\label{Fig:PWinput_k0eqkc_real_imag_abs}Top: Transmission spectrum of a finite sized one dimensional photonic crystal with a cavity. The red dashed line indicates the single mode approximation to the transmission, which is the underlying assumption of the CMT. Bottom: Complex spectrum showing the discrete distribution of QNM resonance frequencies in the complex frequency plane.}
\end{figure}

We start by considering the one dimensional system from Fig.~\ref{Fig:exampleFilter} consisting of a single photonic crystal cavity coupled to free space at both sides. Despite the one dimensional nature, the example is well suited for illustrating the basic principles in CMT, since many systems of interest in integrated optics and related experiments consist of waveguides coupled to cavities in an in-line configuration~\cite{Johnson_OE_14_817_2006,Husko_APL_94_021111_2009} and thus are effectively one dimensional. %
The cavity is formed by two dielectric barriers on either side of a central region. 
The barriers have relative permittivity $\epsilon_\text{r}=13$ and radii $r=0.1a$, where $a$ is the barrier spacing. The background material is air, and the infinite periodic arrangement of barriers in this case results in a photonic crystal with a large band gap~\cite{Joannopoulos2008}. The top panel in Fig.~\ref{Fig:cavity_1D_Q_160_pwTransmission} shows the transmission spectrum for the system. Despite the finite size of the structure, the band gap is clearly visible with a clear resonance at $\omega a/2\pi c\approx0.3$. 
Each of the peaks in the transmission spectrum is related to a specific QNM of the cavity, as can be clearly appreciated when comparing to the complex QNM spectrum in the bottom panel. For the present analysis, we shall focus on the mode for which the real part is inside the band gap and which we denote the cavity mode. The transmission due to this mode alone closely resembles a Lorentzian, as expected for a mode with a finite lifetime. 

\begin{figure}[htb]
\flushright
\begin{overpic}[width=7.2cm]{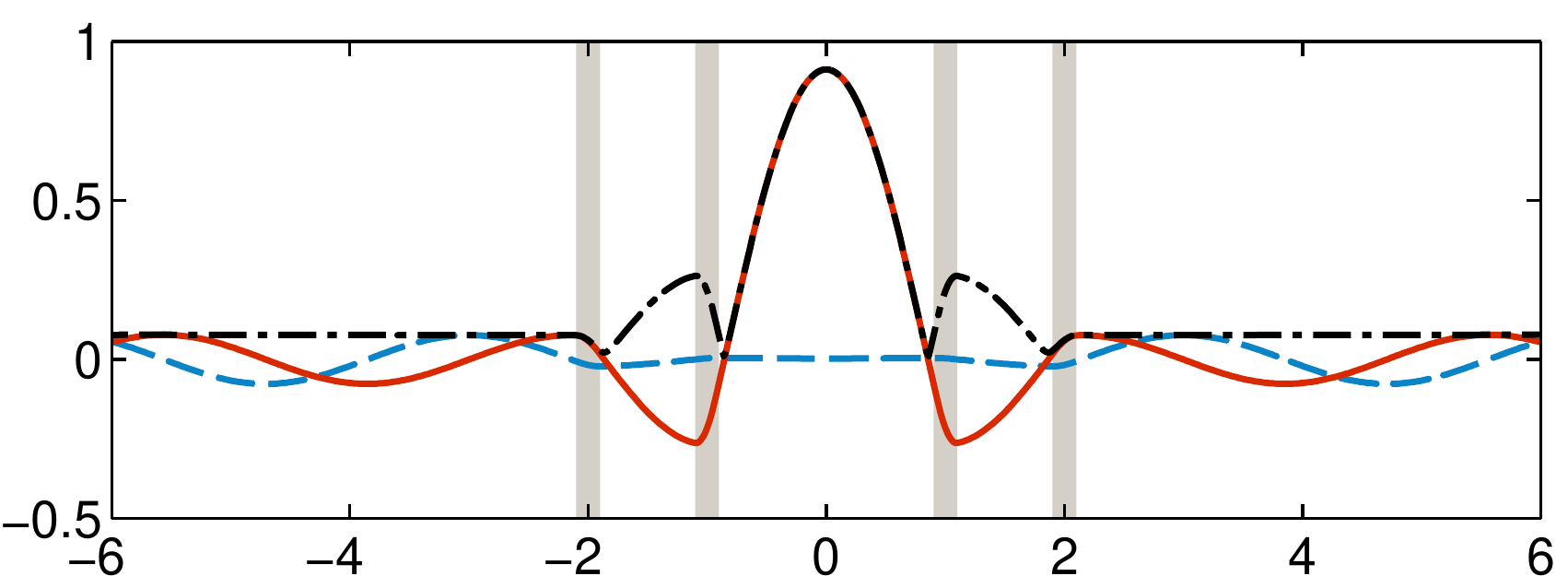}
\put(42,-5){Position $x/a$}
\put(-14,7){\begin{sideways}{Cavity mode}\end{sideways}}
\put(-8,14){\begin{sideways}{$\ft_\text{c}\sqrt{a}$}\end{sideways}} 
\end{overpic}\;\;\;\\[1mm]
\caption{\label{Fig:pcCavity1D_QNM}Real (solid red) and imaginary (dashed blue) parts as well as the absolute value (dashed-dotted black) of the normalized cavity mode of interest in the finite sized one dimensional photonic crystal (indicated by the gray shaded areas). The mode has a complex resonance frequency of $\tlo a/2\pi c=0.2925-0.0009\,\text{i}$, corresponding to $Q=160$.}
\end{figure}

To calculate the cavity mode, we note that the one dimensional version of Eq.~(\ref{Eq:SilverMuller}) may be written as
\begin{align}
\frac{\ud}{\ud x}\ft_\mu(x)\Big\vert_{x=\pm L} = \pm\text{i}n_\text{b}\tilde{k}\ft_\mu(x)
\label{Eq:1DSilverMuller}
\end{align}
where $+/-$ refers to the right/left boundary of the calculation domain of length $2L$. In this case, Eq.~(\ref{Eq:HelmholtzEq}) may be solved to arbitrary accuracy with any standard frequency domain method, \ptk{although the practical implementation is slightly complicated by the fact that the eigenvalue $\tlo$ enters in the boundary condition via $\tilde{k}=\tlo/\text{c}$. One approach is to solve the eigenvalue problem with different trial values $\tlo_\text{guess}$ in the boundary condition and look for cases in which the resulting eigenvalue $\tlo$ is sufficiently close to the trial value. In this way, one can map out the complex QNM spectrum as in the bottom panel of Fig.~\ref{Fig:cavity_1D_Q_160_pwTransmission}, which shows the logarithm of the norm $|\tlo-\tlo_\text{guess}|$, and a dark spot signifies the position of a QNM frequency.} 
For the one dimensional example, the cavity mode of interest, which we denote by $\mu=\text{c}$, has complex resonance frequency $\tlo_\text{c} a/2\pi c=0.292462 - 0.000917\text{i}$, corresponding to $Q\approx160$. This mode is shown in Fig. \ref{Fig:pcCavity1D_QNM}, and 
for positions inside the cavity region it appears to be localized from multiple reflections at the dielectric barriers. The field profile resembles the total field in the one dimensional example in Fig.~\ref{Fig:exampleFilter}, but with the important difference that the field in Fig.~\ref{Fig:pcCavity1D_QNM} is traveling away from the cavity at both sides and 
diverges in the limits $x\rightarrow\pm\infty$, as expected from the boundary condition in Eq.~(\ref{Eq:1DSilverMuller}) with $\text{Im}\{\tilde{k}_\text{c}\}<0$. \ptk{For a translationally invariant waveguide, we can choose the length of the unit cell $a$ to be arbitrarily small. In the limit of small unit cell, the QNM norm in Eq.~(\ref{Eq:QNM_Norm}) reduces to the well established norm for QNMs in one dimension~\cite{Lai_PRA_41_5187_1990, Muljarov_EPL_92_50010_2010}}.

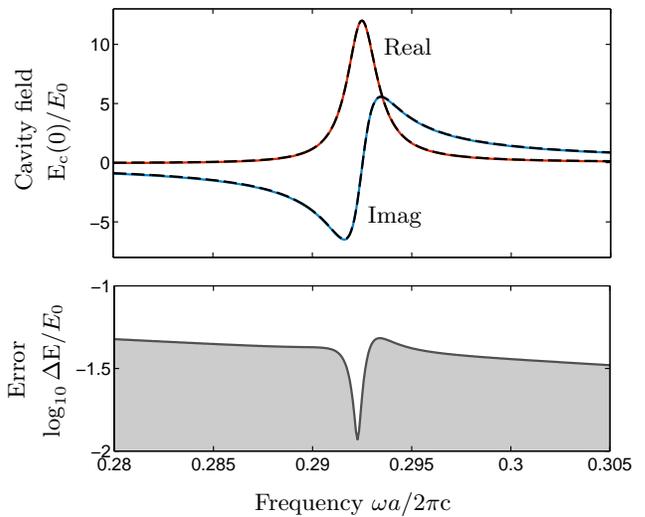
\begin{figure}[htb]
\flushright
\begin{overpic}[width=7cm]{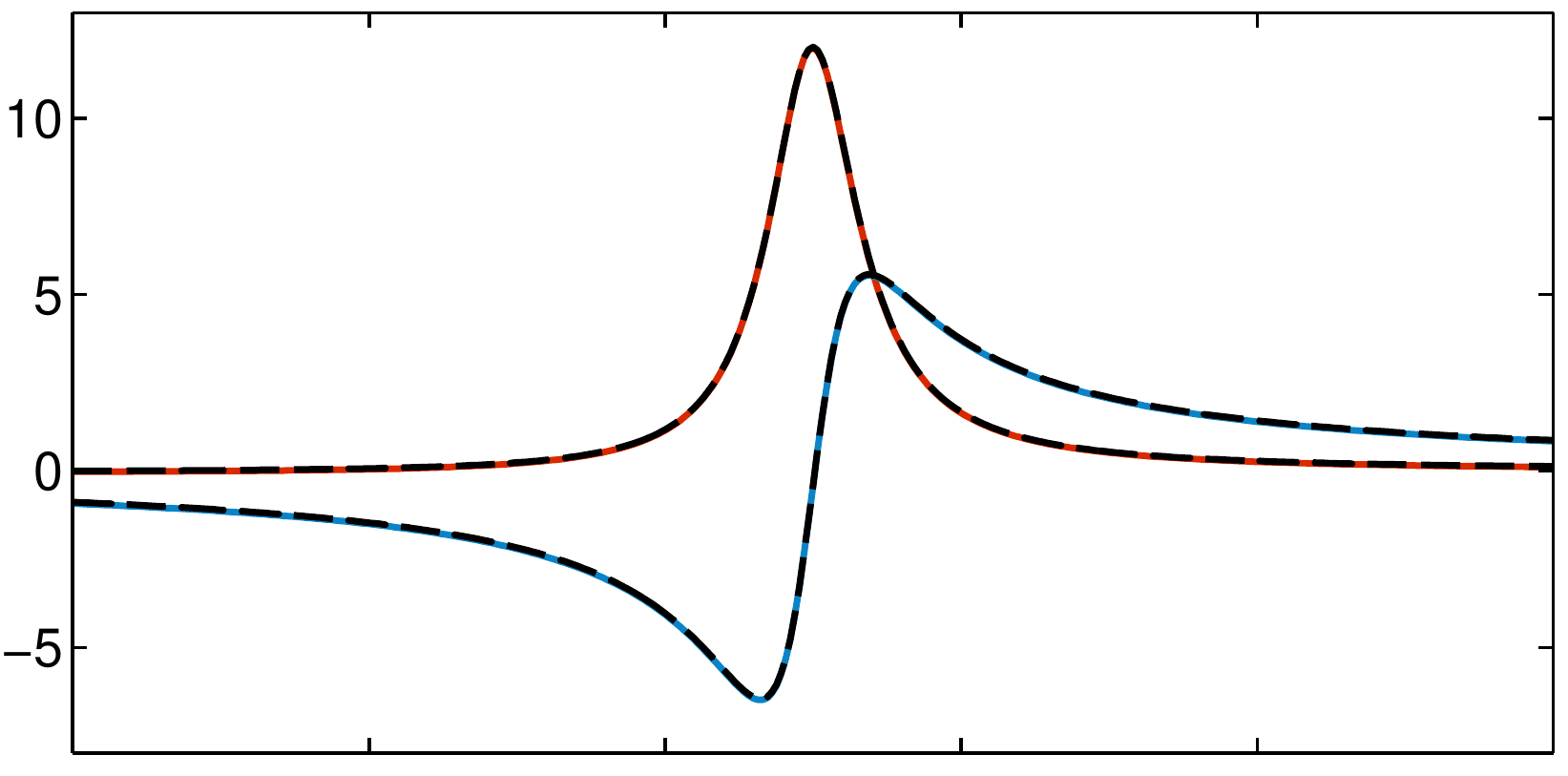}
\put(56,40){Real}
\put(53,8){Imag}
\put(32,-7){Frequency $\omega a/2\pi\text{c}$}
\put(-14,14){\begin{sideways}{Cavity field}\end{sideways}}
\put(-8,16){\begin{sideways}{$\text{E}_\text{c}(0)/E_0$}\end{sideways}}
\end{overpic}\;\;\;\;\\[1mm]
\begin{overpic}[width=7.45cm]{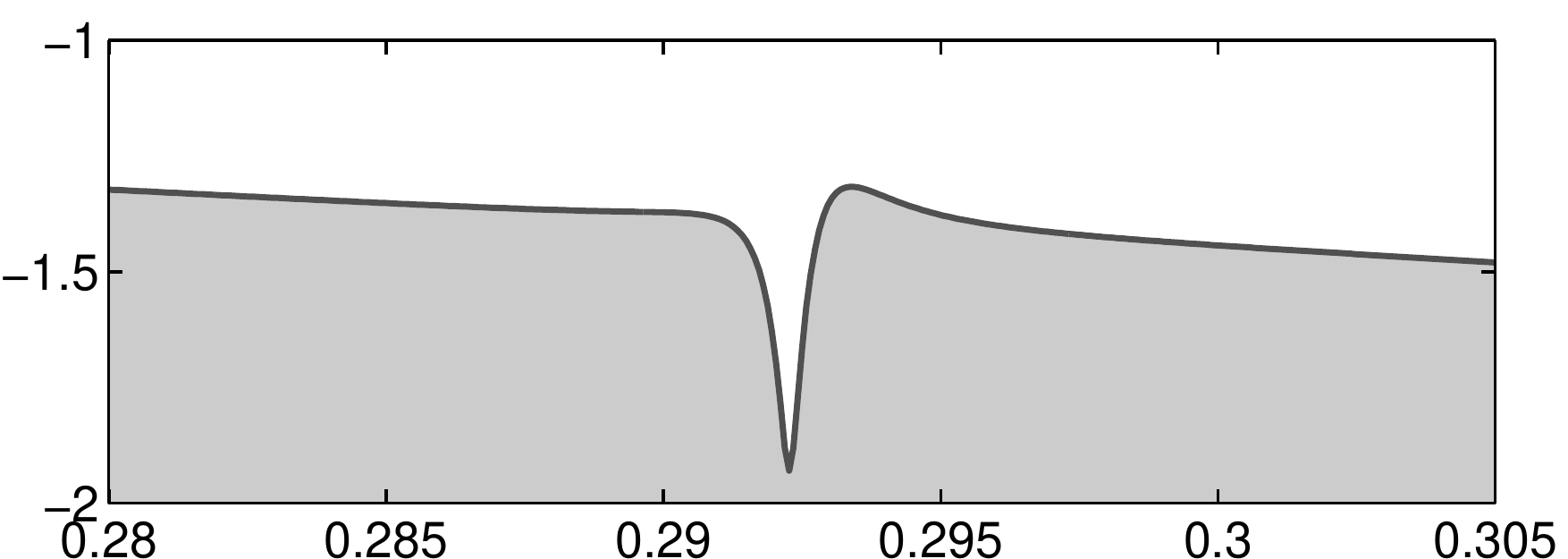}
\put(32,-7){Frequency $\omega a/2\pi\text{c}$}
\put(-12,13){\begin{sideways}{Error}\end{sideways}}
\put(-6,6){\begin{sideways}{$\log_{10}\Delta \text{E}/E_0$}\end{sideways}}
\end{overpic}\;\\[5mm]
\caption{\label{Fig:pcCavity1D_coupling}Top: Real and imaginary parts of the complex field in the cavity center when illuminated by a plane wave from the left. Dashed black curves show the CMT result $\text{E}_\text{c}^\text{CMT}(0,\omega)=\Gamma(\omega)E_{1+}\ft_\text{c}(0)$. Bottom: Error $\Delta \text{E}=|\text{E}^\text{CMT}_\text{c}(0,\omega)-\text{E}^\text{ref}_\text{c}(0,\omega)|$ between the two calculations in the top panel. 
}
\end{figure}

In one dimension, the general waveguide modes are simply plane waves of the form $\text{f}_{n\pm}(x,t) =\exp\{\pm\text{i}kx\}$ with $x<-2.1a$ $(n=1)$ or $x>2.1a$ $(n=2)$, and we choose the phase of the waveguide modes so that they are purely real at the onset of the waveguides at $x=\mp2.1a$. 
In this case, the parameters of interest are listed in Table~\ref{Tab:pcCavity_1D_parameters}. 
\begin{table}[htb]
\centering
\begin{tabular}{lccc}
 Parameter & Notation & Value & Units\\
\hline
Resonance frequency  & $\tlo_\text{c} $ & 0.292462 - 0.000917i &  $2\pi\text{c}/a$ \\
Coupling  & $\sigma_{1\text{c}}$ & 0.075812 - 0.005831i & $1/\sqrt{a}$\\ 
Group velocity & $v_\text{g}$ & 1& $\text{c}$ \\
Center value & $\ft_\text{c}(0)$ &  0.912555 + 0.002628i & $1/\sqrt{a}$\\
\end{tabular}
\caption{\label{Tab:pcCavity_1D_parameters}CMT parameters for the coupled waveguide-cavity-waveguide system in Fig.~\ref{Fig:pcCavity1D_QNM}. The coupling constants are calculated with the phase convention that the waveguide modes are purely real at $x=\pm2.1a$.
}
\end{table}
To assess the coupling in Eq.~(\ref{Eq:C_def}), we solve the full scattering problem numerically and use this as a reference. Figure~\ref{Fig:pcCavity1D_coupling} shows a comparison between the result from Eq.~(\ref{Eq:E_cav_from_E_wg}) and the reference calculation. The lower panel shows the absolute error, which is less than 0.05 throughout the relatively large bandwidth shown; at resonance it drops to approximately 0.01, corresponding to a relative error of less than one part in a thousand. 

\subsubsection*{Time domain behavior}
\begin{figure}[htb]
\flushright
\begin{overpic}[width=7.525cm]{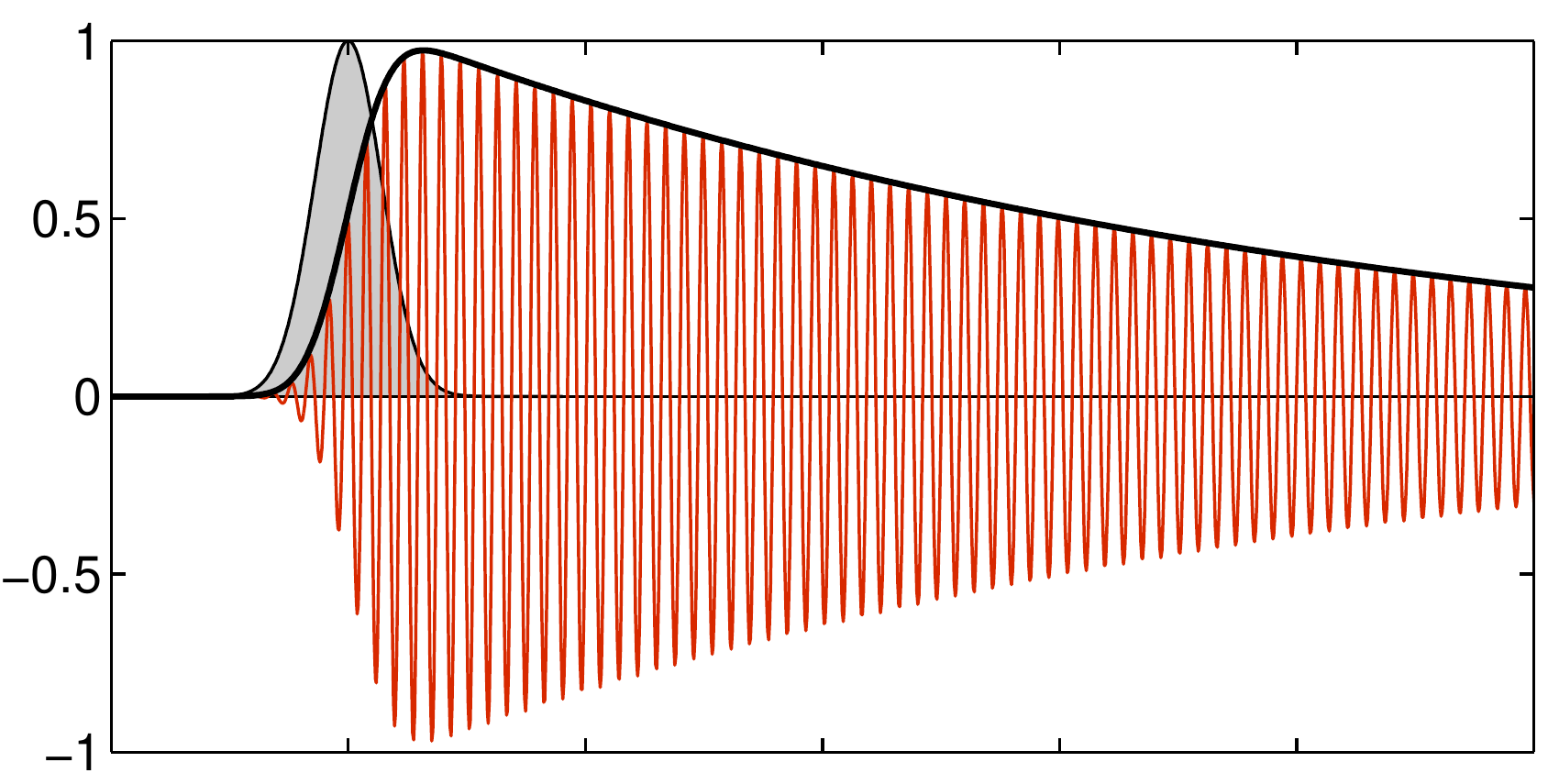}
\put(9,41.5){(a)}
\put(-13,13){\begin{sideways}{Electric field}\end{sideways}}
\put(-6,1){\begin{sideways}{$\text{E}_\text{c}(0)/E_0$ or $\text{E}_{1+}(x_\text{D})/E_0$}\end{sideways}}
\end{overpic}\;\;\\[.8mm]
\begin{overpic}[width=7.5cm]{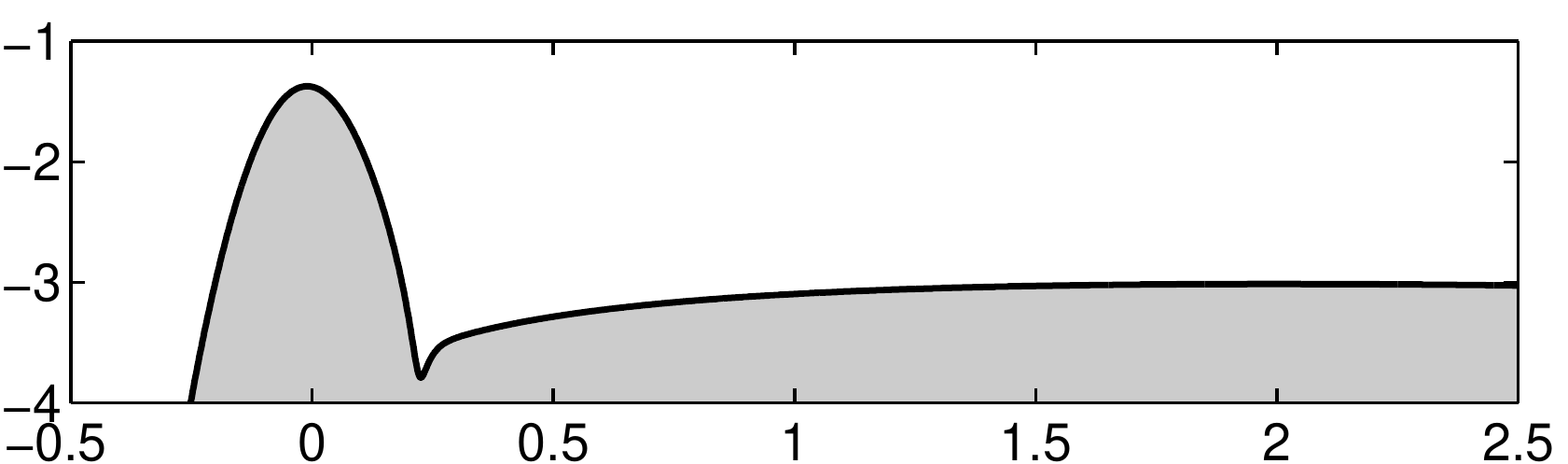}
\put(-15.75,10){\begin{sideways}{Error}\end{sideways}}
\put(-8.75,3){\begin{sideways}{$\log_{10}\Delta \text{E}/E_0$}\end{sideways}}
\put(6.5,21){(b)}
\put(41.5,-5){Time, $2\gamma_\text{c} t$}
\end{overpic}\\[6mm]
\begin{overpic}[width=7.7cm]{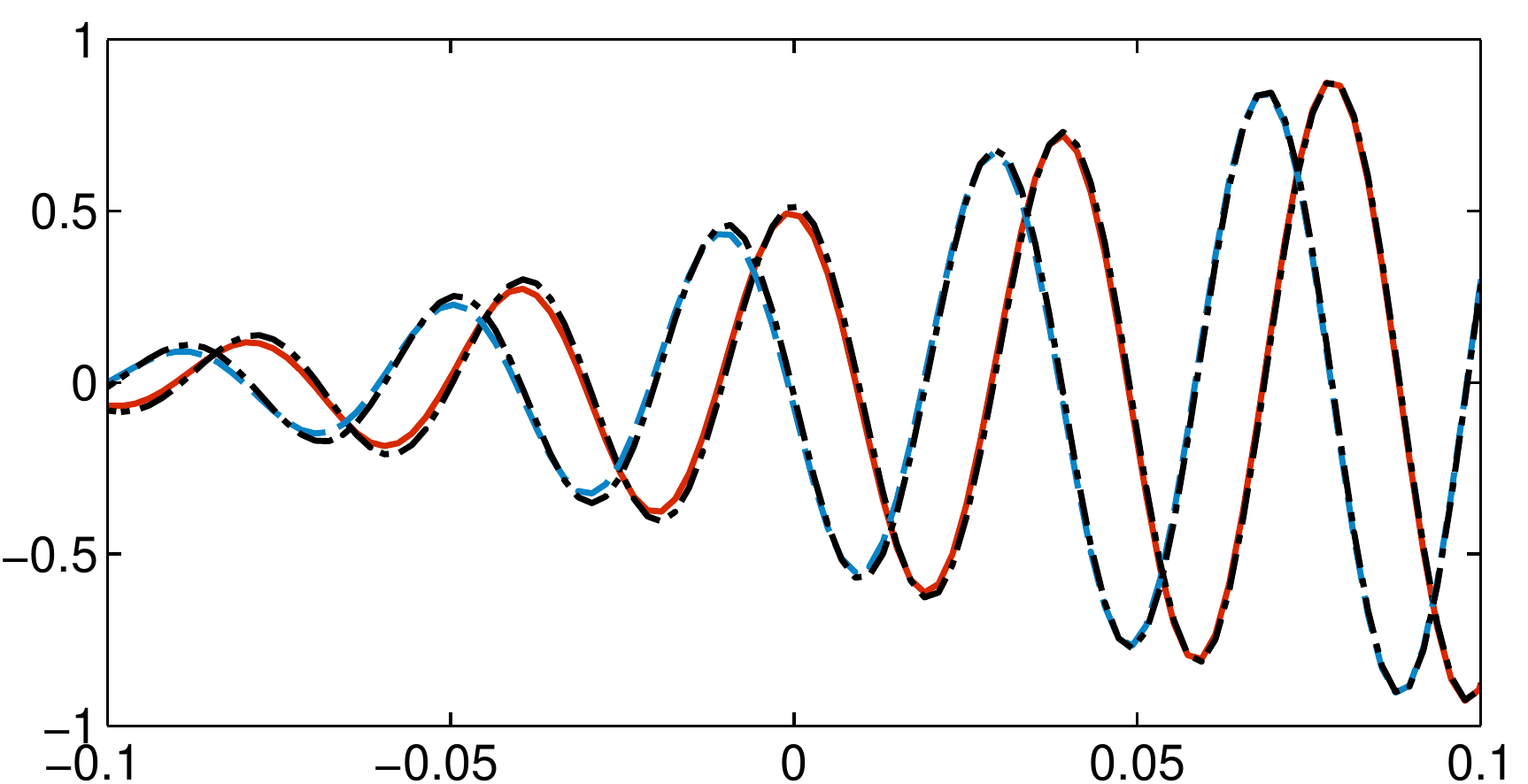}
\put(9,43){(c)}
\put(-9,5.5){\begin{sideways}{Cavity field, $\text{E}_\text{c}(0)/E_0$}\end{sideways}}
\put(42.5,-5){Time, $2\gamma_\text{c} t$}
\end{overpic}\\[6mm]
%
%
\caption{\label{Fig:cavityWGresponse_cavField_Q_160_linear_response}(a): Electric field in the center of the cavity as a function of time. Thin red curve shows the total field of the reference calculation, 
and full black curve shows the absolute value of the field as calculated by CMT in Eq.~(\ref{Eq:single_mode_cavity_response_gaussian}). Gray shading indicates the electric field strength of the input pulse at the edge of the cavity. (b): Error $\Delta \text{E}=|\text{E}^\text{CMT}_\text{c}(0,t)-\text{E}^\text{ref}_\text{c}(t)|$ in the single-mode CMT calculation. 
(c): Zoom-in of the solutions at around $2\gamma t=0$ showing the real (red solid) and imaginary (blue dashed) parts of the total field in the cavity center. Black dashed-dotted curves show the CMT result.}
\end{figure}
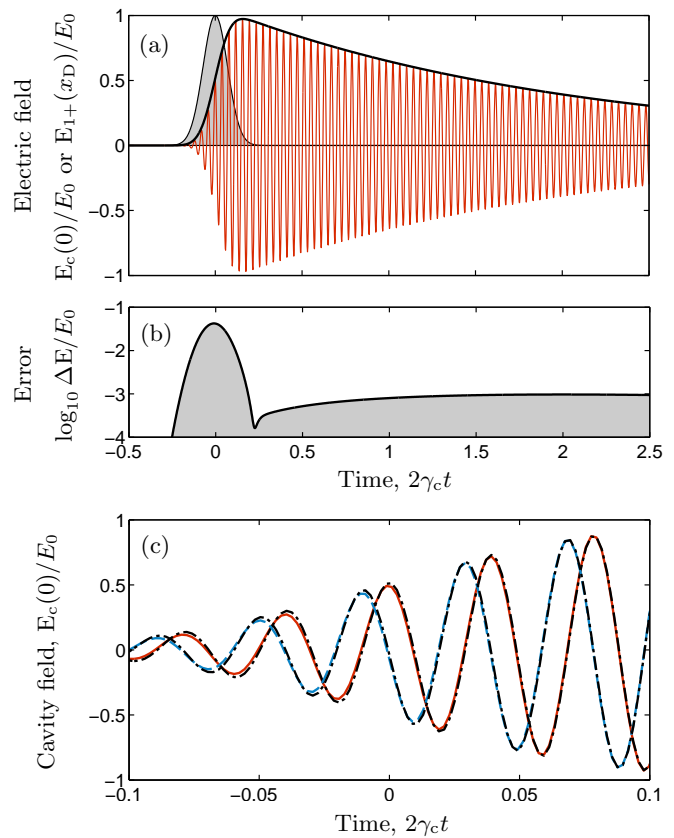

We consider again the one dimensional example in Fig.~\ref{Fig:exampleFilter}, 
but consider now the time-domain dynamics of the field in the center of the cavity.  
We choose a Gaussian input pulse of the form
\begin{align}
E_{1+}(x_\text{D},t) = E_0 \text{e}^{-\left(\omega_\text{c} t/s\right)^{2}}\text{e}^{-\text{i}\omega_\text{c} t},
\label{Eq:gaussianInput}
\end{align}
where $x_\text{D}=-2.1a$ and $s=Q/10$, and the carrier frequency of the input pulse is resonant with the cavity, $\omega_\text{c}=\text{Re}\{\tlo_\text{c}\}$.
With this input, the single resonance response in Eq.~(\ref{Eq:temporal_field_as_fourier_intergral}) can be calculated analytically and is given as
\begin{align}
E^\text{CMT}_\text{c}(t) = E_0\frac{\sqrt{\pi}s v_\text{g}\sigma}{2\omega_\text{c}}\text{e}^{-\text{i}\tlo_\text{c}t}\text{e}^{\xi^2}\left[1-\text{Erf}\left(\xi-\frac{\omega_\text{c} t}{s}\right)\right],
\label{Eq:single_mode_cavity_response_gaussian}
\end{align} 
where $\xi=\gamma_\text{c}s/2\omega_\text{c}$. From Fig.~\ref{Fig:cavity_1D_Q_160_pwTransmission}, however, it is clear that there may be contributions from other modes to the dynamics. Indeed, we consider these contributions to be the source of the errors in the frequency domain comparison in Fig.~\ref{Fig:pcCavity1D_coupling}. For the reference calculation, therefore, we use the full numerical Fourier transform of $E_\text{c}(0,\omega)$. %
%
%
Panel (a) of Fig.~\ref{Fig:cavityWGresponse_cavField_Q_160_linear_response} shows the time-dependent electric field in the cavity center. 
In addition, we show the magnitudes of the input field and the cavity field when calculated using the CMT solution in Eq.~(\ref{Eq:single_mode_cavity_response_gaussian}). The latter represents the output from typical CMT calculations based on Eq.~(\ref{Eq:linearFieldEq}) in which the phase of the field is left unspecified. In contrast, Eq.~(\ref{Eq:single_mode_cavity_response_gaussian}) holds the full information about the cavity field, including the phase. Panel (b) shows the absolute error in the single-mode CMT calculation, which is largest throughout the duration of the input pulse and drops to approximately one part in a thousand at later times when the field is left to evolve freely. Panel (c) shows a zoom-in of the solution to highlight the small disagreement between the two solutions at around $t=0$ when the deviation is most pronounced.

\subsection{Side coupled cavities in two dimensions}
\label{Sec:Example_2D}
We now turn to the slightly more complicated case of side coupled cavities in a PC made from dielectric cylinders in a square lattice with spacing $a$ . The rods have radius $r=0.2a$ and relative permittivity $\epsilon_\text{cyl}=8.9$, and the background is assumed to be air. Before moving on to the double cavity structure in Fig.~\ref{Fig:exampleFilter}, we start by considering the case of a single side coupled cavity at the distance $d=2a$ from the center of the waveguide~\cite{Kristensen_OL_39_6359_2014, deLasson_OL_40_5790_2015}. The cavity supports a cavity mode at the complex frequency $\tlo_\text{c}=0.39687 - 0.00136\text{i}$, corresponding to $Q=146$~\cite{Kristensen_OL_39_6359_2014}. 

\begin{figure}[htb]
\flushright
\begin{overpic}[width=7.38cm]{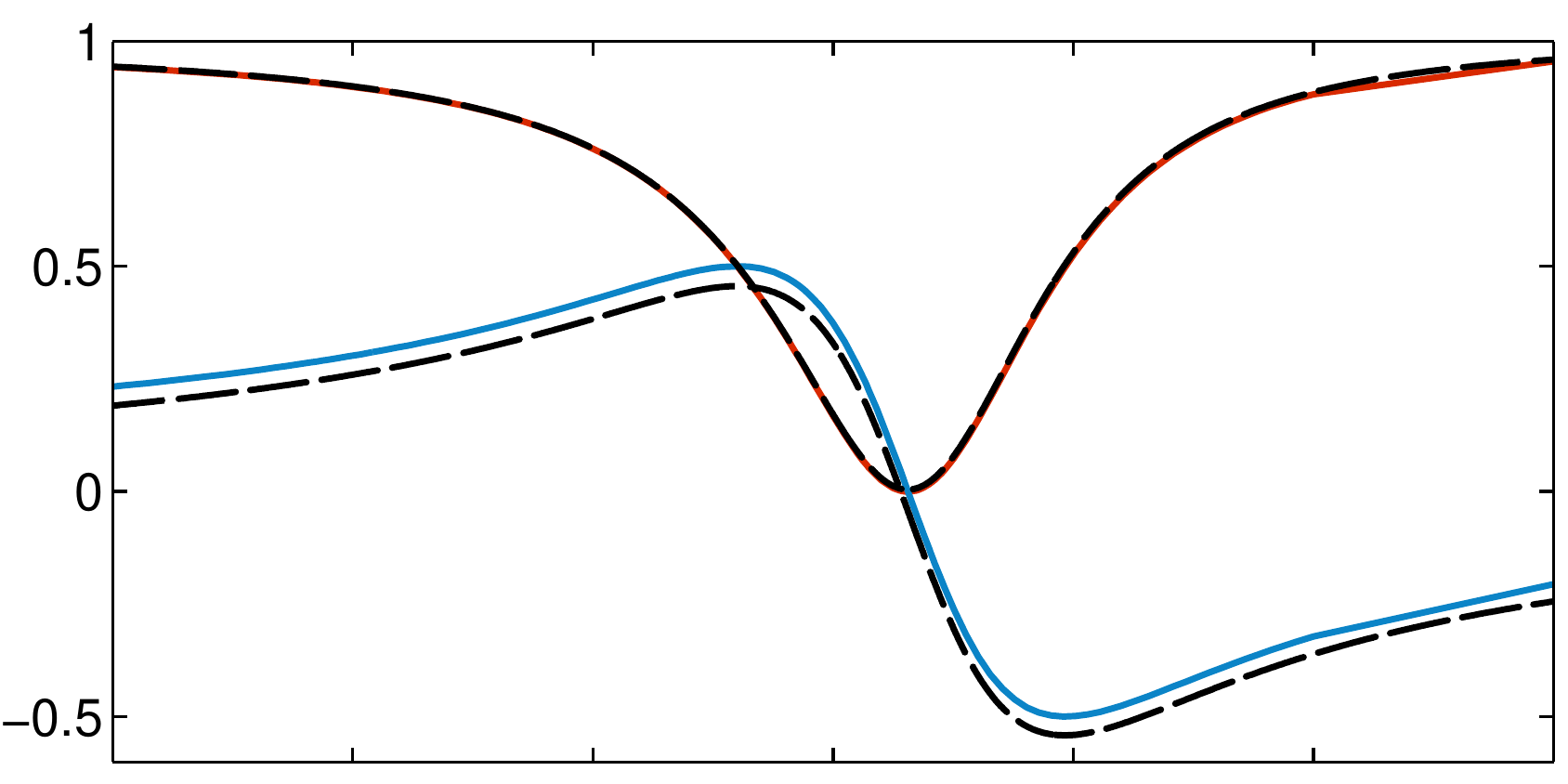}
\put(87,38){Real}
\put(87,15){Imag}
\put(32,-7){Frequency $\omega a/2\pi\text{c}$}
\put(-8,10){\begin{sideways}{Transmission, $T$}\end{sideways}}
\end{overpic}\;\;\;\\[1mm]
\begin{overpic}[width=7.45cm]{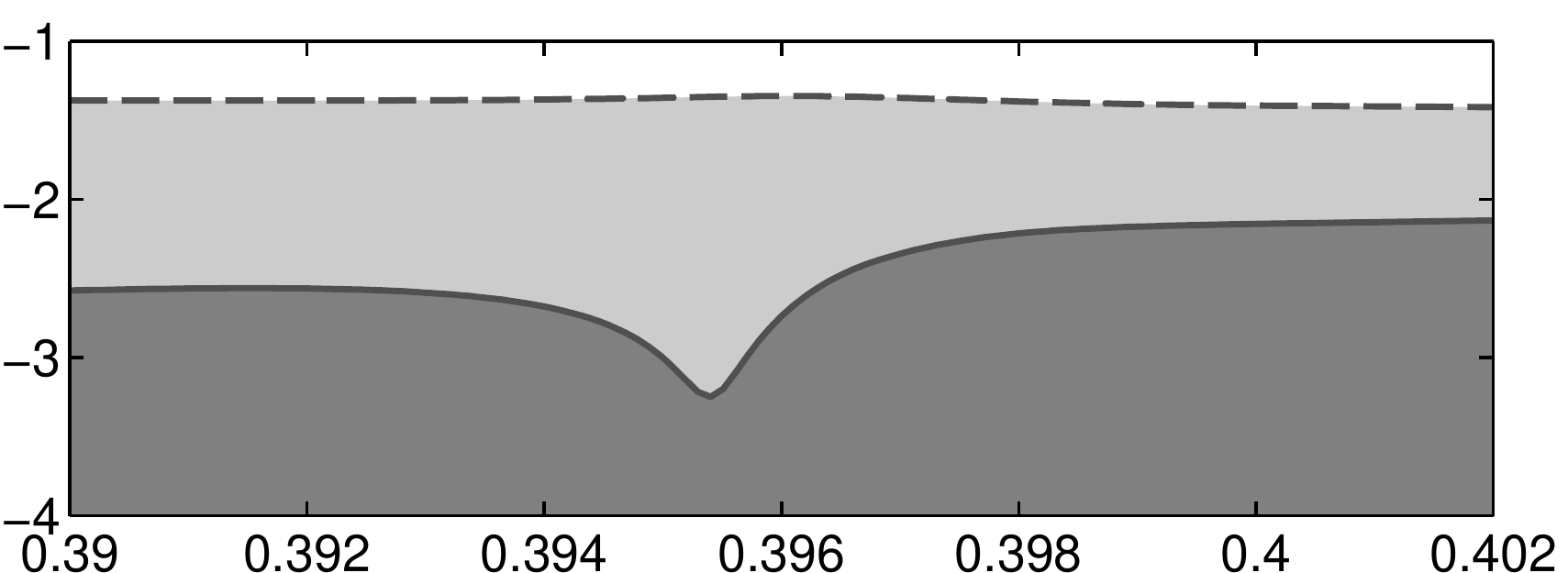}
\put(32,-7){Frequency $\omega a/2\pi\text{c}$}
\put(-14,14){\begin{sideways}{Error}\end{sideways}}
\put(-8,11){\begin{sideways}{$\log_{10}\Delta T$}\end{sideways}}
\end{overpic}\\[5mm]
\caption{\label{Fig:TransRefl_dCav2_dEpsVar_w_FMM_phase_0p05_no_propPhase}Top: Real and imaginary parts of the complex transmission through the PC waveguide with a single side coupled cavity. 
Dashed black curves show the CMT result when setting the phase $\Phi=1$ in Eq.~(\ref{Eq:sideCoupledTrans}). Bottom: Error $\Delta T=|T^\text{CMT}_\text{c}(\omega)-T^\text{ref}_\text{c}(\omega)|$ between the two calculations in the top panel using $\Phi=1$ (light gray and dashed curve) and $\Phi=\exp\{0.05\text{i}\}$ (dark gray and solid curve).
}
\end{figure}
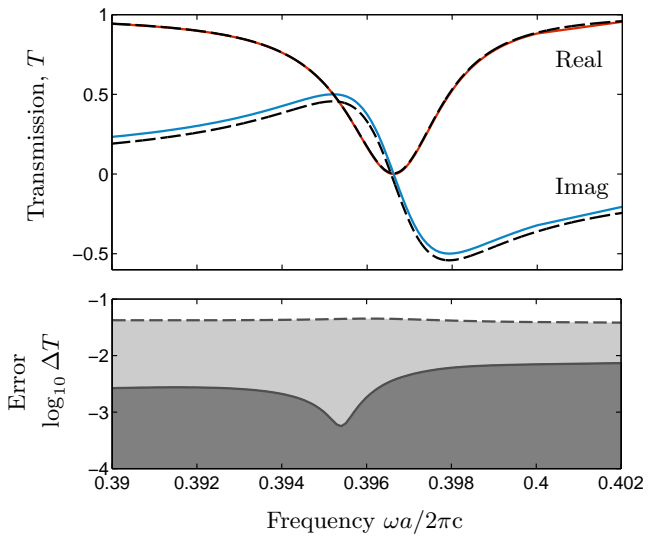

The waveguide supports forward and backwards propagating modes $\mf_+(\mr)$ and $\mf_-(\mr)$, for which the phase is chosen so that the modes are purely real in the plane through the center of the cavity. The QNM of interest is symmetric with respect to this plane~\cite{Kristensen_OL_39_6359_2014}, wherefore the complex coupling to the two waveguide modes is identical, $\sigma_+=\sigma_-=\sigma_\text{c}$. As an approximation for the Green tensor, we use the sum of the waveguide Green tensor in Eq.~(\ref{Eq:GreensTensorWG}) and the single-term QNM expansion of the Green tensor in the cavity from Eq.~(\ref{Eq:GreensTensorCavity}). For use in transmission calculations, we focus on the case $x>x'$, for which we can write the Green tensor as
\begin{align}
\mG(\mr,\mr',\omega) &\approx \frac{\text{i}}{2k}\frac{\text{c}}{v_\text{g}}\mf_+(\mr)\mf_-(\mr')\Phi + \frac{\mft_\text{c}(\mr)\mft_\text{c}(\mr')}{2k(\tlk_\text{c}-k)},
\label{Eq:G_trans_with_Phi}
\end{align}
where $\Phi$ is a general phase resulting from the influence of the cavity on the field in the waveguide and therefore is expected to tend to the limiting case of $\Phi=1$ when the cavity is far from the waveguide. The QNMs in the cavity can be related to the forward and backward propagating modes in the single waveguide via the coupling parameters $\sigma_\pm$, so that the transmission becomes
\begin{align}
T(\omega) = \Phi+\text{i}\frac{v_\text{g}}{\omega-\tlo_\text{c}}\sigma_+\sigma_- = \Phi+\Gamma_\text{c}(\omega)\sigma_\text{c}^2.
\label{Eq:sideCoupledTrans}
\end{align}
To assess the transmission in Eq.~(\ref{Eq:sideCoupledTrans}) we solve the full wave propagation problem numerically by means of the Fourier modal method (FMM) employing a Bloch mode expansion and S-matrix technique~\cite{deLasson_PhD_2015, Lecamp_OE_15_11042_2007, Gregersen_2014}. This method is well suited for transmission calculations, since it gives immediate access to the transmission between the guided modes on either side of the cavity. We use the same numerical framework to calculate the QNM in the cavity~\cite{deLasson_JOSA_A_31_2142_2014} and list the parameters of importance in setting up the CMT in Table~\ref{Tab:pcCavity_2D_parameters}. All calculations were performed using 101 Fourier components and 128 staircasing steps in each unit cell, which explains the slight deviation between the resonance frequency in Table~\ref{Tab:pcCavity_2D_parameters} and the one reported in Ref.~\cite{Kristensen_OL_39_6359_2014}.
\begin{table}[htb]
\centering
\begin{tabular}{lccc}
Parameter & Notation & Value & Units \\
\hline
Resonance frequency  & $\tlo_\text{c} $ & $0.39668 - 0.00136\text{i}$ &  $2\pi\text{c}/a$ \\
Coupling  & $\sigma_\text{c}$ & 0.00544 - 0.12753i & $1/\sqrt{a}$ \\
Group velocity & $v_\text{g}$ &  0.52294& $\text{c}$ \\
\end{tabular}
\caption{\label{Tab:pcCavity_2D_parameters}CMT parameters for a PC waveguide with a single side coupled cavity.
}
\end{table}
Figure~\ref{Fig:TransRefl_dCav2_dEpsVar_w_FMM_phase_0p05_no_propPhase} shows the comparison between the CMT result and the reference calculations. Using the simple choice of $\Phi=1$, which we expect to be valid in the case of high $Q$-values, we find a relatively good agreement between the reference calculation and the CMT. The simple theory clearly captures the qualitative behavior of the transmission and has a maximum absolute error on the order of 0.05. Increasing the phase slightly improves the agreement, and for $\Phi=\exp\{0.05\text{i}\}$ the absolute error is on the order of 0.001 on resonance, as shown in the bottom panel of Fig~\ref{Fig:TransRefl_dCav2_dEpsVar_w_FMM_phase_0p05_no_propPhase}. The transmission through a side coupled cavity shows a minimum around the cavity mode frequency, which arises from interference of the two terms in Eq.~(\ref{Eq:sideCoupledTrans}). The interference makes the calculations very sensitive to the discretization, but by using the same discretization for the QNM calculation and the reference calculations, we expect any residual error due to discretization to be the same in the two calculations; numerical investigations with varying discretization confirm this. For this reason, we attribute the small error to the inherent approximate nature of the single mode expansion in the CMT.

\subsubsection*{Two side coupled cavities}
\begin{figure}[b]
\flushright
\vspace{-2mm}
\begin{overpic}[width=3.5cm]{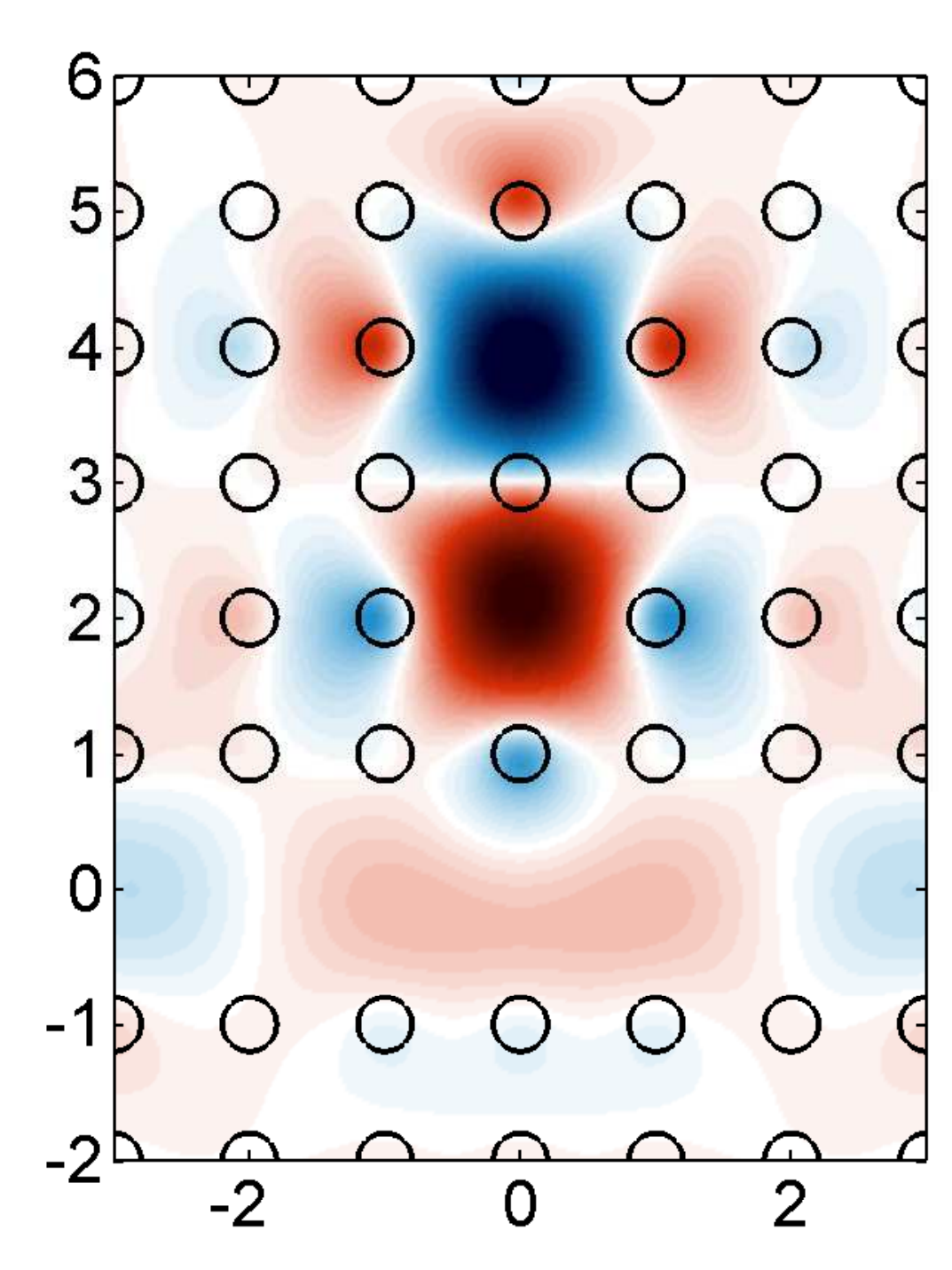}
\put(22,-7){Position $x/a$}
\put(-8,31){\begin{sideways}{Position $y/a$}\end{sideways}}
\end{overpic}
\begin{overpic}[width=4.56cm]{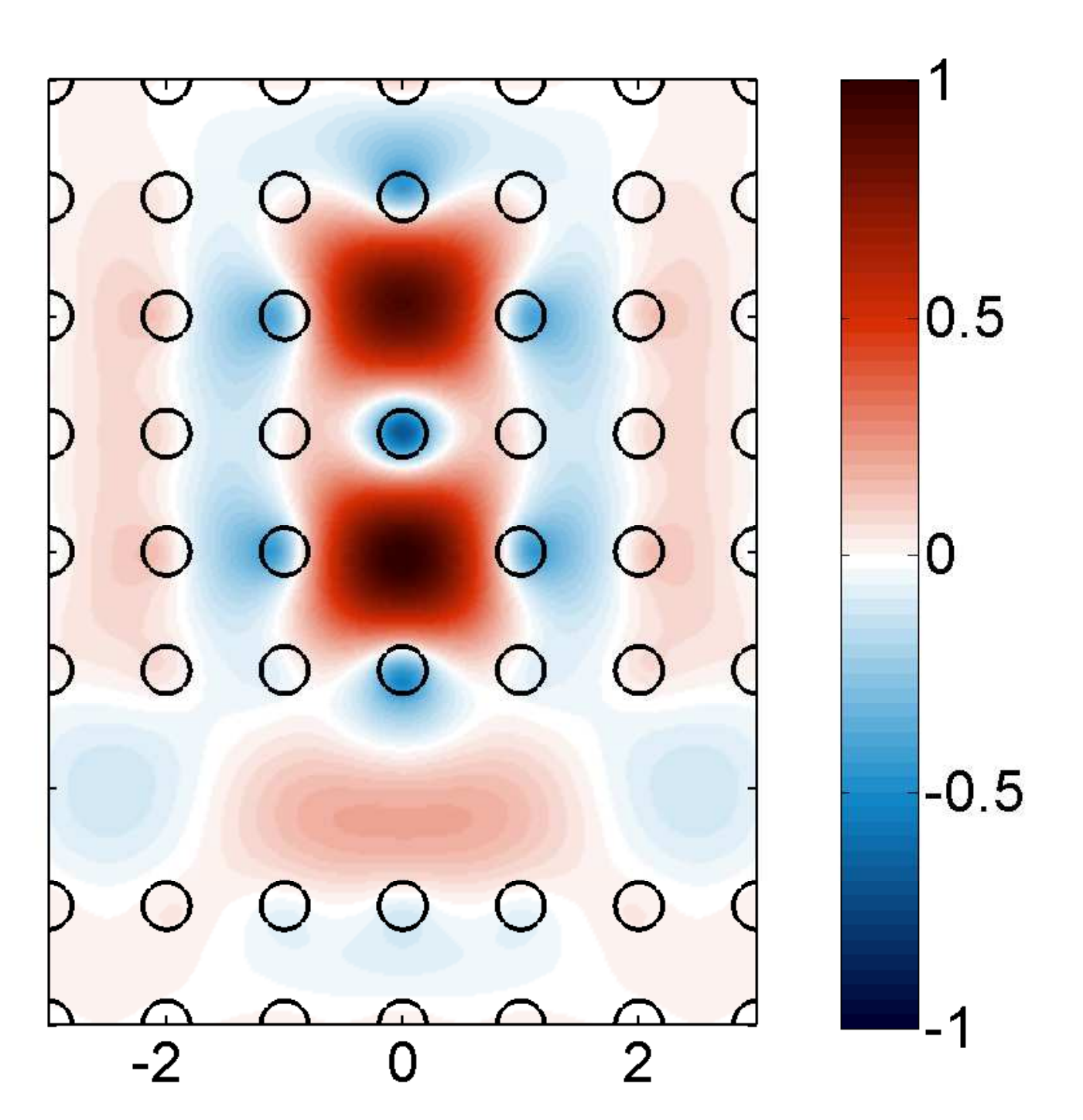}
\put(16,-7){Position $x/a$}
\end{overpic}\\[2mm]
\caption{\label{Fig:QNMs_DoubleCav}Mode profiles showing the real part of the QNMs of interest with complex resonance frequencies $\tlo_1=0.3850 - 0.0008\text{i}$ (left) and $\tlo_2 = 0.4058 - 0.0006\text{i}$ (right). The QNMs are scaled to unity in the center of the cavity closest to the waveguide.\\[-4mm]}
\end{figure}

As a last example, we return to the double cavity structure in Fig.~\ref{Fig:exampleFilter} to illustrate how the QNM based approach to the CMT works in this case of a  more complicated system. 
The procedure is almost identical to that of the single side coupled cavity, the only difference being an additional QNM in the bandwidth of interest. Figure~\ref{Fig:QNMs_DoubleCav} shows the two QNMs with field distributions which appear even and odd with respect to the plane separating the cavities, as expected from the limiting case of coupled cavities in an infinite PC with no waveguide. Considering the symmetry with respect to the plane through the centers of the cavities, however, both modes are symmetric, and therefore the complex couplings to the two waveguide modes are identical as in the case of the single side coupled cavity.
Following an approach similar to the single cavity case, we approximate the Green tensor as the sum of the waveguide Green tensor and a QNM expansion of the Green tensor in the cavities as
\begin{align}
\mG(\mr,\mr',\omega) &\approx \frac{\text{i}}{2k}\frac{\text{c}}{v_\text{g}}\mf_+(\mr)\mf_-(\mr')\Phi + \sum_\mu\frac{\mft_\mu(\mr)\mft_\mu(\mr')}{2k(\tlk_\mu-k)},
\label{Eq:G_trans_with_Phi_and_sum}
\end{align}
in which the sum runs over the two modes of interest, $\mu=1$ or $\mu=2$. The CMT approximation to the transmission then becomes
\begin{align}
T(\omega) = \Phi+\Gamma_1(\omega)\sigma_1^2++\Gamma_2(\omega)\sigma_2^2.
\label{Eq:sideCoupledTrans_doubleCav}
\end{align}

Because of the strong similarity between this system and that of the single side coupled cavity, we expect the influence of the cavities on the field in the waveguide to be the same. Therefore, this example also serves to justify the introduction of the phenomenological extra phase $\Phi$ since, to a first approximation, we expect this extra phase to be independent of the extra cavity. 

\begin{figure}[htb]
\flushright
\begin{overpic}[width=7.58cm]{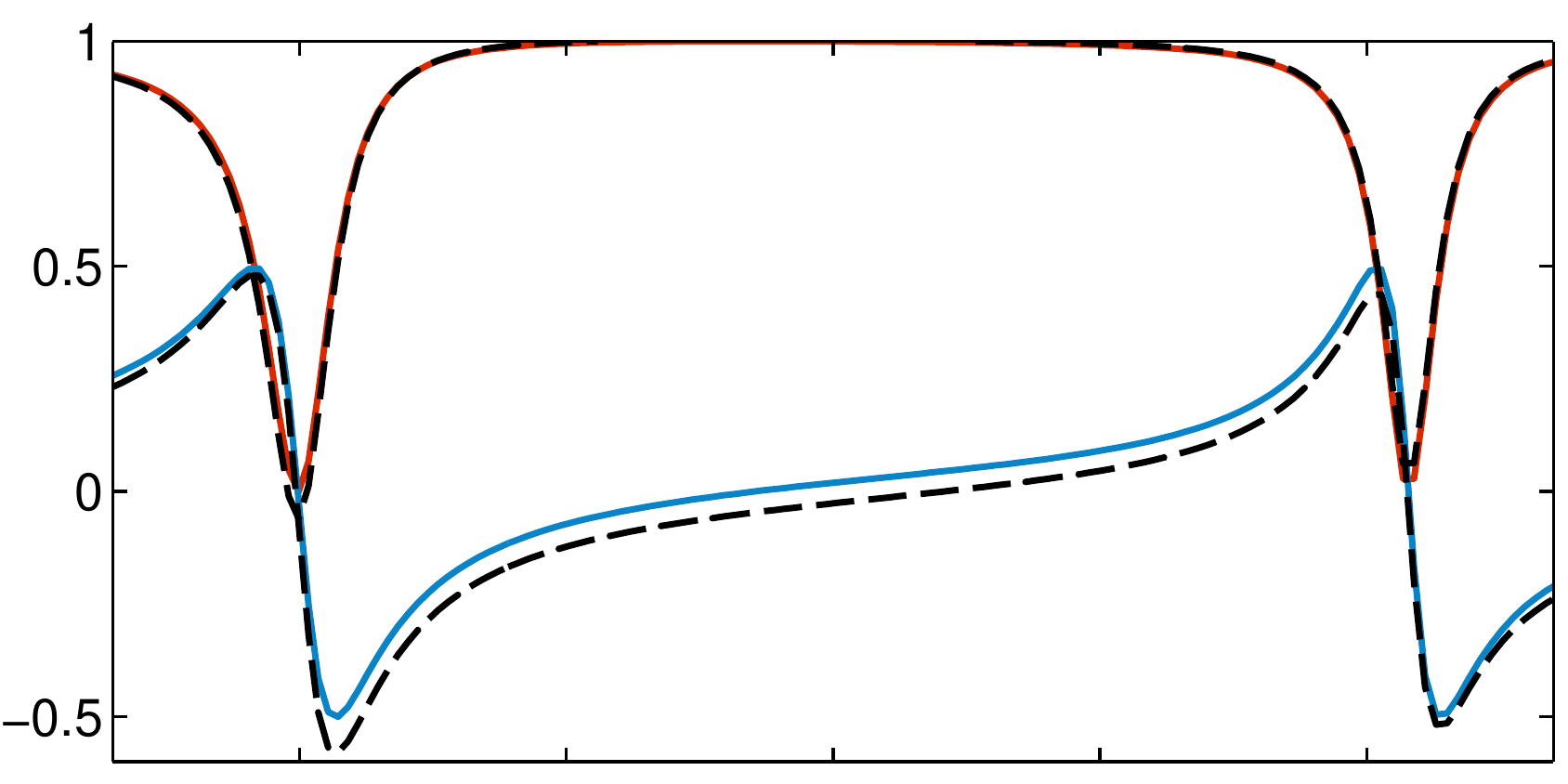}
\put(71,41){Real}
\put(78,7){Imag}
\put(32,-7){Frequency $\omega a/2\pi\text{c}$}
\put(-8,10){\begin{sideways}{Transmission, $T$}\end{sideways}}
\end{overpic}\;\;\\[1mm]
\begin{overpic}[width=7.35cm]{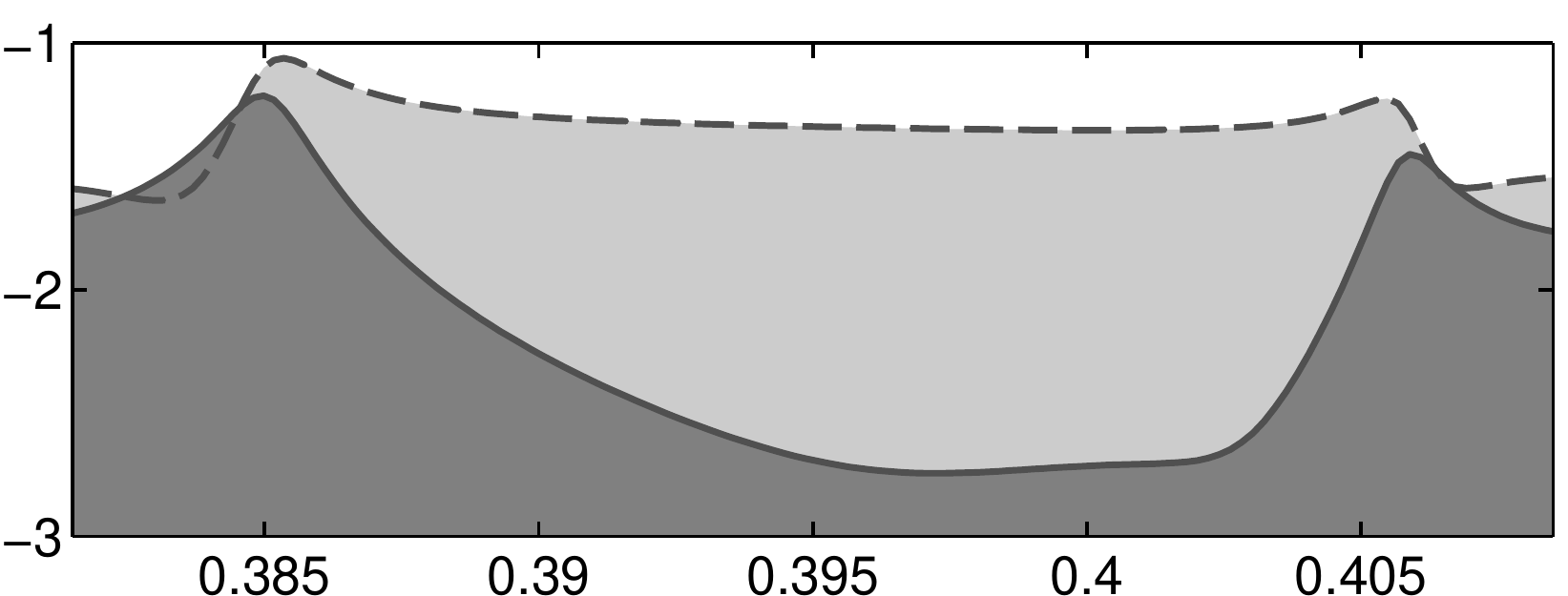}
\put(32,-7){Frequency $\omega a/2\pi\text{c}$}
\put(-14,14){\begin{sideways}{Error}\end{sideways}}
\put(-8,11){\begin{sideways}{$\log_{10}\Delta T$}\end{sideways}}
\end{overpic}\;\;\\[5mm]
\caption{\label{Fig:TransRefl_dCav2_dEpsVar_w_FEM_phase_0p05_no_propPhase}Top: Real and imaginary parts of the complex transmission through the coupled cavity-waveguide system with two cavities. Dashed black curves show the CMT result when setting the phase $\Phi=1$ in Eq.~(\ref{Eq:sideCoupledTrans}). Bottom: Error $\Delta T=|T^\text{CMT}_\text{c}(\omega)-T^\text{ref}_\text{c}(\omega)|$ between the two calculations in the top panel using $\Phi=1$ (light gray and dashed curve) and $\Phi=\exp\{0.05\text{i}\}$ (dark gray and solid curve).
}
\end{figure}
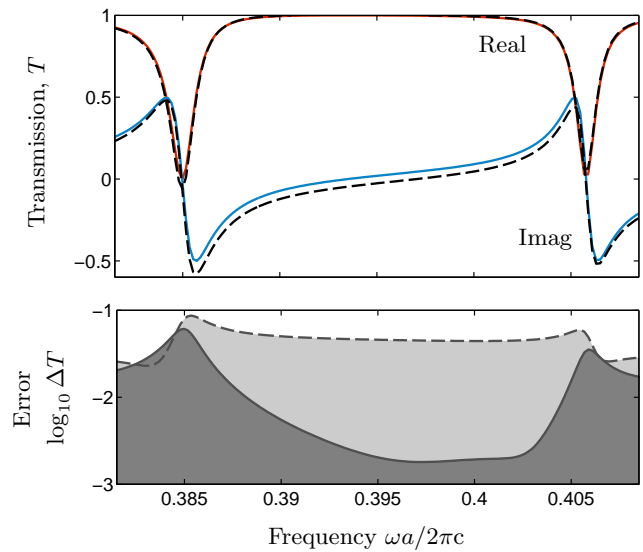

The parameters of importance for setting up the CMT model are listed in Table~\ref{Tab:pcCavity_2D_DoubleCav_parameters}, and the full complex transmission is shown in Fig.~\ref{Fig:TransRefl_dCav2_dEpsVar_w_FEM_phase_0p05_no_propPhase} for both the cases of $\Phi=1$ and $\Phi=\exp\{0.05\text{i}\}$.
\begin{table}[htb]
\centering
\begin{tabular}{lccc}
Parameter & Notation & Value & Units \\
\hline
Resonance frequency  & $\tlo_1 $ & $0.38502 - 0.00075\text{i}$ &  $2\pi\text{c}/a$ \\
Resonance frequency  & $\tlo_2 $ & $0.40578 - 0.00060\text{i}$ &  $2\pi\text{c}/a$ \\
Coupling  & $\sigma_1$ & 0.00711 - 0.09779\text{i} & $1/\sqrt{a}$ \\
Coupling  & $\sigma_2$ & 0.00009 - 0.08357\text{i} & $1/\sqrt{a}$ \\
Group velocity & $v_\text{g}$ &  0.52294& $\text{c}$ \\
\end{tabular}
\caption{\label{Tab:pcCavity_2D_DoubleCav_parameters}CMT parameters for the structure in Fig.~\ref{Fig:QNMs_DoubleCav} with two side coupled cavities.
}
\end{table}
As in the previous case, we compare to independent reference calculations performed with the same FMM code and using the same numerical settings, whereby we argue that the observed differences can be attributed to the inherent approximate nature of the CMT. Using $\Phi=1$, the maximum errors are approximately twice as large as in the single cavity case, and increasing the phase lowers the error dramatically to a few parts in a thousand at the single cavity resonance frequency. The overall agreement appears less impressive than in the single cavity case, with maximum errors on the order $0.06$, which we attribute to the larger bandwidth and the larger complexity of the material system. Nevertheless, for many research or design applications, this may be a small price to pay for the enormous simplification and physical insight offered by the CMT approach.



\section{Conclusion}
\label{Sec:Conclusion}
We have presented an alternative derivation of the (temporal) coupled mode theory (CMT) for modeling of light propagation in systems of optical cavities coupled to waveguides. We have argued, that the cavity modes can be naturally modeled as quasinormal modes (QNMs) with complex resonance frequencies corresponding to a finite lifetime due to the field leaking from the cavities. The leaky nature of the QNMs make them distinctly different from the waveguide modes, which propagate through the waveguide without decay. We have discussed how one can use the field equivalence principle to couple the two families of modes in a frequency domain description of the coupled system. By transforming to the time domain, one can recover the well known CMT description, Eqs.~(\ref{Eq:linearFieldEq}) and (\ref{Eq:wglOutput}), in the limit of high cavity $Q$-values. The theory itself, however, is not limited to cavities with high $Q$-values, and we have assessed the accuracy of the method by comparing to explicit reference calculations in one and two dimensions with moderate $Q$-values. The relative errors in both cases were found to be as low as 0.001, when allowing for a phenomenological phase in the case of the side coupled cavities.

This alternative derivation shows  that the cavity modes in CMT can be explicitly defined as the QNMs that leak through the waveguides (as well as other channels in general), thus removing much of the ambiguity surrounding the calculation and normalization of these modes. From a practical modeling point of view, we expect the theory to be useful in modeling and design of optical experiments and devices for which details of the phase relation between different transmission channels are of importance as in the case of two side coupled cavities, for example. Also, the definition of the cavity modes as QNMs provides a framework for setting up more advanced models of cavity-enhanced non-linear dynamics and general light-matter interaction in a precise and unambiguous way. Last, we remark that the use of the field equivalence principle is not restricted to QNMs in cavities coupled to waveguides, and we expect it to be useful also in other material systems for modeling the coupling of a general incoming field to the QNMs of optical or plasmonic resonators.

\appendix
\section{Expanding QNMs on waveguide modes}
\label{App:Sigma_def}
Expansion of the QNMs in terms of the analytical continuation of the waveguide modes 
in Eq.~(\ref{Eq:sigmaTildeDef}) becomes particularly clear in a modal picture~\cite{Lecamp_OE_15_11042_2007, Gregersen_2014, deLasson_PhD_2015}, where, for positions in the waveguide sections, the electric field QNMs $\mft_\mu(\mr)$ and the magnetic field QNMs $\tilde{\mathbf{g}}_\mu(\mr)$ are expanded on the full set of Bloch modes traveling away from the cavity, including propagating, evanescent and non-guided modes as~\cite{deLasson_JOSA_A_31_2142_2014}
\begin{align}
\mft_\mu(\mr) &= \sum_n \sigma_{\mu n}\mf_{n-}(\mr,\tlo_\mu)
	\label{Eq:sigmaTildeDef_G} \\
		\tilde{\mathbf{g}}_\mu(\mr) &= \sum_m \sigma_{\mu m}\mathbf{g}_{m-}(\mr,\tlo_\mu).
		\label{Eq:sigmaTildeDef_G_H}	
\end{align}
The Bloch modes are computed at the complex frequency $\tlo_\mu$, and $\sigma_{\mu n}$ are complex expansion coefficients. In a modal picture, therefore, the $\sigma_{\mu n}$ corresponding to the guided modes can be obtained directly from the scattering matrices describing coupling of Bloch modes in the various periodic elements. In the general case, they may be calculated by suitable projections of the QNMs onto the analytical continuation of the waveguide modes by exploiting the orthogonality relation~\cite{Lecamp_OE_15_11042_2007}
\begin{align}
\int_A \big[\mathbf{f}_{m-} \times \mathbf{g}_{n+} - \mathbf{f}_{n+} \times \mathbf{g}_{m-} \big] \cdot \mathbf{n}\, \ud A = \delta_{mn} N_n,
\end{align}
where $N_n$ is a normalization constant. Here, and in Eq.~(\ref{Eq:sigma_from_integral}) below, we have suppressed the explicit position and frequency dependence of the modes. Formation of the difference $\mft_\mu(\mr) \times \mathbf{g}_{n+}(\mr,\tlo) - \mathbf{f}_{n+}(\mr,\tlo) \times \tilde{\mathbf{g}}_\mu(\mr) $ and integration over the plane $D$ perpendicular to the wave\-guide then leads to
\begin{align}
\sigma_{\mu n} = \frac{\int_D \big[\mft_\mu\times\mathbf{g}_{n+} - \mathbf{f}_{n+} \times \tilde{\mathbf{g}}_\mu \big] \cdot \mathbf{n}\, \ud A }{\int_D \big[\mathbf{f}_{n-} \times \mathbf{g}_{n+} - \mathbf{f}_{n+} \times \mathbf{g}_{n-} \big] \cdot \mathbf{n}\, \ud A },
\label{Eq:sigma_from_integral}
\end{align}
where all modes are evaluated at $\tlo_\mu$. 

\section{The coupling parameter $\sigma_{\mu n}$}
\label{Sec:CouplingParameter}
A key assumption in the application of the field equivalence principle for the derivation of the CMT, is that the cavity modes can be related to the waveguide modes at the real part of the cavity resonance frequency as in Eq.~(\ref{Eq:sigmaDef}). To zero'th order, this relation follows from the definition of the waveguide radiation condition in Eq.~(\ref{Eq:sigmaTildeDef}), whereby, at positions sufficiently far from the cavity in waveguide $n$, the QNM $\mft_\mu(\mr)$ can be written via analytical continuation in terms of the waveguide mode $\mf_{n-}(\mr,\tlo_\mu)$ traveling away from the cavity. Because both sets of modes are normalized, the expansion coefficient $\sigma_{\mu n}$ is in general a non-trivial complex number, but for any fixed choice of modes, $\sigma_{\mu n}$ is well defined. From Eq.~(\ref{Eq:sigmaTildeDef}), with $\tlo_\mu=\omega_\mu-\text{i}\gamma_\mu$, and expanding around $\tlo_\mu=\omega_\mu$ we find
\begin{align}
\mft_\mu(\mr) &\approx
\sigma_{\mu n}\mf_{n-}(\mr,\omega_\mu) -\text{i}\sigma_{\mu n}\gamma_\mu\partial_\omega\mf_{n-}(\mr,\omega_\mu).
\label{Eq:fmu_sigma_zero_and_first_order}
\end{align}
Using the Bloch form of the waveguide modes in Eq.~(\ref{Eq:Blochform}) we can investigate the second term 
by rewriting the derivative as
\begin{align}
\partial_\omega \mf_{n-}(\mr,\omega_\mu) &= \frac{1}{v_\text{g}}\partial_k\mf_{n-}(\mr,\omega_\mu) \\
&=\frac{1}{v_\text{g}}\text{e}^{\text{i}\mk\cdot\mr} \Big[\text{i}\,[\me_\mk\cdot\mr]\,\mathbf{u}_\mk(\mr) + \partial_k\mathbf{u}_\mk(\mr) \Big],
\end{align}
where $\me_\mk$ denotes a unit vector in the direction of the waveguide. In this way, we can rewrite Eq.~(\ref{Eq:fmu_sigma_zero_and_first_order}) as a position dependent expansion onto a waveguide mode at real frequencies and a correction term as
\begin{align}
\mft_\mu(\mr) \approx\;  &\sigma_{\mu n}\left[1 + \frac{\gamma_\mu}{v_\text{g}}[\me_\mk\cdot\mr]\right]\mf_{n-}(\mr,\omega_\mu)\nonumber \\ &-\text{i}\frac{\sigma_{\mu n}\gamma_\mu}{v_\text{g}}\text{e}^{\text{i}\mk\cdot\mr}\partial_k\mathbf{u}_\mk(\mr).
\label{Eq:sigma_linearCorrect}
\end{align}
The waveguide modes are determined only up to an arbitrary phase factor that we may write as $\exp\{-\text{i}\mk\cdot\mr_D\}$, where $\mr_D$ is in the plane $D$. Thus, the second term in the square brackets can be regarded as small for $|x-x_D|\ll v_\text{g}/\gamma_\mu$. Since the end goal is to describe the QNMs in terms of the waveguide modes, we generally choose the plane to be as close to the cavity section as possible, yet still within the (possibly discrete) translationally invariant part of the geometry defining the waveguide. This distinction between the different parts of the geometry is directly built in to the FMM.


\section{Use of the Dyson equation}
\label{App:Dyson}
The Dyson equation provides the Green tensor of a general structure defined by the relative permittivity $\epsilon_\text{r}(\mr)$ in terms of the known Green tensor $\mG_\text{B}(\mr,\mr',\omega)$ of a reference structure defined by $\epsilon_\text{B}(\mr)$ as
\begin{align}
&\mG(\mr,\mr',\omega) = \mG_\text{B}(\mr,\mr',\omega) \nonumber \\
&\qquad+ k^2\int_V\mG_\text{B}(\mr,\mr'',\omega)\Delta\epsilon(\mr'')\mG(\mr'',\mr',\omega)\ud V,
\end{align}
in which $\Delta\epsilon_\text{r}(\mr) = \epsilon_\text{r}(\mr)-\epsilon_\text{B}(\mr)$ denotes the difference in relative permittivity between the two structures. To apply the Dyson equation to set up a relatively simple model for a general coupled cavity-waveguide structure, we can consider the change $\Delta\epsilon_\text{r}(\mr)$ to be the change in permittivity defining the optical cavity in an otherwise (possibly discrete) translationally invariant background for which we can easily calculate or estimate the Green tensor. With such a choice, $\Delta\epsilon_\text{r}(\mr)$, and hence the integral, will be non-zero only inside the cavity, where we expect the expansion of the Green tensor on one or just a few QNMs to be adequate. Therefore, we write
\begin{align}
&\mG(\mr,\mr',\omega) \approx \mG_\text{B}(\mr,\mr',\omega) \nonumber \\
&\qquad+ k^2\int_V\mG_\text{B}(\mr,\mr'',\omega)\Delta\epsilon(\mr'')\sum_\mu\frac{\mft_\mu(\mr'')\mft_\mu(\mr')}{2k(\tlk_\mu-k)}\ud V.
\end{align}
Next, we expand the background Green tensor on the waveguide modes of the reference structure, and assuming the waveguide modes to be approximately zero at positions within the cavity, we have $\mG_\text{B}(\mr,\mr',\omega)\approx0$, so
\begin{align}
\mG(\mr,\mr',\omega) \approx \sum_\mu\frac{\mft_\mu(\mr)\mft_\mu(\mr')}{2k(\tlk_\mu-k)},
\end{align}
for $\mr$ and $\mr'$ both in the cavity. This, in turn, reduces the overall approximation of the Green tensor to the physically appealing form in Eq.~(\ref{Eq:G_overall_approx}).

\section{Energy conservation}
\label{App:Energy}
For a single mode cavity, the total power $P$ leaking from the cavity can be split in different channels $P_{n-}$ as
\begin{align}
P = P_{1-} + P_{2-} + ... = -2[\gamma_1+\gamma_2+...]\langle U\rangle,
\end{align}
where $\langle U\rangle$ denotes the average energy in the cavity, and the sign convention is such that the power is positive for energy going into the cavity. For a cavity with electric field given as $\mE_\mu = E_\mu\mft_\mu(\mr)$ and a decay channel corresponding to waveguide $n$, we may write the power as
\begin{align}
P_{n-} &= \frac{1}{2\mu_0}\int_{D_n} \text{Re}\left\{\mE_\mu(\mr)\times\mB_\mu^*(\mr) \right\}\cdot\mathbf{n}\,\ud A \\
&\approx \frac{1}{2\mu_0}E_\mu^2\int_{D_n} \text{Re}\bigg\{ \sigma_{\mu n}\mf_{n-}(\mr)\nonumber \\
&\qquad\qquad\quad \times\bigg[\frac{\text{i}}{\omega}\nabla\times\left[\sigma_{\mu n}^*\mf_{n-}^*(\mr) \right]\bigg] \bigg\}\cdot\mathbf{n}\,\ud A \\
&=-|\sigma_{\mu n}|^2 \frac{E_\mu^2}{E_{n+}^2}P_{n+},
\end{align}
where $P_{n+}$ is the average power carried by an incoming electromagnetic field of the form $\mE_\text{in}=E_{n+}\mf_{n+}(\mr,\omega)$, cf. Eqs.~(\ref{Eq:inputWGfield}) and (\ref{Eq:inputWGfield_B}). Therefore, since $P_{n-} =-2\gamma_n\langle U\rangle$ and $P_{1+} = v_\text{g}\epsilon_0E_{n+}^2/2$, we find that
\begin{align}
|\sigma_{\mu n}|^2 = \frac{4\gamma_n\langle U\rangle}{v_\text{g}\epsilon_0E_\mu^2}.
\label{Eq:sigma_mu_n_v1}
\end{align}
For a high-$Q$ cavity, one can choose the phase of the QNMs so that the fields are almost entirely real and so that
\begin{align}
\langle U\rangle = \frac{1}{2}\epsilon_0\int\epsilon_\text{r}(\mr)\mE_\mu(\mr)\cdot\mE_\mu^*(\mr)\,\ud V \approx\frac{1}{2}\epsilon_0E_\mu^2\langle\langle\mft_\mu|\mft_\mu\rangle\rangle,
\end{align}
and, since the QNMs are assumed to be normalized, we can then express the average energy in the cavity as $\langle U\rangle\approx E_\mu^2\epsilon_0/2$. Inserting in Eq.~(\ref{Eq:sigma_mu_n_v1}) we find that
\begin{align}
|\sigma_{\mu n}|^2 \approx 2\frac{\gamma_n}{v_\text{g}},
\label{Eq:sigma_mu_n_approx}
\end{align}
so that the norm of the expansion coefficient $\sigma_{\mu n}$ is the ratio of the rate of energy leakage through the waveguide to the group velocity in the waveguide. In the case of a symmetric two-port cavity we have simply $|\sigma_{\mu n}|^2 = \gamma/v_\text{g}$.



\vspace{5cm}

\end{document}